\documentclass{revtex4}

\usepackage{amsthm,amssymb,amsmath}				
\usepackage{graphicx,comment}
\usepackage{float}   
\usepackage{xcolor}

\begin{document}

\title{Fine tuning of rainbow gravity functions and Klein-Gordon particles in cosmic string rainbow gravity spacetime}
\author{Omar Mustafa}
\email{omar.mustafa@emu.edu.tr}
\affiliation{Department of Physics, Eastern Mediterranean University, G. Magusa, north
Cyprus, Mersin 10 - Turkiye.}

\begin{abstract}
\textbf{Abstract:}\ We argue that, as long as relativistic quantum particles are in point, the variable $y=E/E_p$ of the rainbow functions pair $g_{_{0}} (y)$ and $g_{_{1}} (y)$ should be fine tuned into $y=|E|/E_p$, where $E_p$ is the Planck's energy scale. Otherwise,  the rainbow functions will be only successful to describe the rainbow gravity effect on relativistic quantum particles and the anti-particles will be left unfortunate. Under such fine tuning, we consider Klein-Gordon (KG) particles in cosmic string rainbow gravity spacetime in a non-uniform magnetic field (i.e., $\mathbf{B}=\mathbf{\nabla }\times \mathbf{A}=\frac{3}{2}B_{\circ }r\,\hat{z}$ ).  Then we consider KG-particles in cosmic string rainbow gravity spacetime in a uniform magnetic field (i.e., $\mathbf{B}=\mathbf{\nabla }\times \mathbf{A}=\frac{1}{2}B_{\circ }\,\hat{z}$ ). Whilst the former effectively yields KG-oscillators, the later effectively yields  KG-Coulombic particles.  We report on the effects of rainbow gravity on both KG-oscillators and Coulombic particles using four pairs of rainbow functions: 
(i) $%
g_{_{0}}\left( y\right) =1$, $g_{_{1}}\left( y\right) =\sqrt{1-\epsilon y^{2}%
}$, (ii) $g_{_{0}}\left( y\right) =1$, $g_{_{1}}\left( y\right) =\sqrt{%
1-\epsilon y}$, (iii) $g_{_{0}}\left( y\right) =g_{_{1}}\left( y\right)
=\left( 1-\epsilon y\right) ^{-1}$, and (iv) $g_{_{0}}\left( y\right) =\left(
e^{\epsilon y}-1\right) /\epsilon y$, $g_{_{1}}\left( y\right) =1$, where $y=|E|/E_p$ and $\epsilon$ is the rainbow parameter. It is interesting to report that, all KG particles' and anti-particles' energies are symmetric about $E=0$ value (a natural relativistic quantum mechanical tendency), the invariance of the Planck's energy scale is only observed for the family of  rainbow functions used in loop quantum gravity (i.e., the pairs in (i) and (ii)), whereas the energies for pairs (iii) and (iv) fail to show any eminent convergence towards the Planck's energy scale $E_p$, and a phenomenon of energy states to fly away and disappear from the spectrum is observed for the rainbow functions pair (iii) at $\gamma=\epsilon m/E_p=1$.


\end{abstract}

\maketitle

\section{Introduction}

Quantum gravity (QG), a semi-classical model of rainbow gravity (RG),
has attracted research attention over the years \cite{R1,R2,R3,R4,R5}. Under the RG-model, the energy of the probe particles is assumed to affect the spacetime background, at the ultra-high energy
regime, so that the spacetime metric become an energy-dependent one \cite{R5,R6,R7,R8,R81,R9,R10,R11,R12}. Hereby, the Planck energy $E_{p}=\sqrt{\hbar
c^{5}/G}$  plays the role of a threshold separating the classical description from the quantum mechanical one and introduces itself as another invariant energy scale alongside the speed of light. Consequently, rainbow
gravity justifies the modified relativistic energy-momentum dispersion relation%
\begin{equation}
E^{2}g_{_{0}}\left( y\right) ^{2}-p^{2}c^{2}g_{_{1}}\left( y\right)
^{2}=m^{2}c^{4};\; y=E/E_{p},  \label{e2}
\end{equation}%
where $g_{_{0}}\left( y\right) $, $g_{_{1}}\left( y\right) $ are the rainbow functions, $E$ is the energy of the probe particle and $mc^{2}$ is its rest
mass energy. Such a modification in the energy-momentum relation is significant in the ultraviolet limit and is constrained to reproduce the standard GR dispersion relation in the infrared limit so that 
\begin{equation}
\lim\limits_{y\rightarrow 0}g_{_{k}}\left( y\right) =1;\;k=0,1.  \label{e3}
\end{equation}%
The effects of such modifications could be observed, for example, in the tests of thresholds for ultra high-energy cosmic rays \cite{R6,R61,R8,R81,R13,R14,R15}, TeV photons \cite{R16}, gamma-ray bursts \cite{R6,R61,R8}, nuclear physics experiments \cite{R17}.

In the rainbow gravity settings, recent studies on the quantum mechanical gravity effects are carried out. Amongst are, the thermodynamical properties of black holes \cite{R18,R19,R20,R21,R211}, the dynamical stability conditions of neutron stars \cite{R22}, thermodynamic stability of modified black holes \cite{R221}, charged black holes in massive RG \cite{R222}, on geometrical thermodynamics and heat engine of black holes in RG \cite{R223}, on RG and f(R) theories \cite{R224}, the initial singularity problem for closed rainbow cosmology \cite{R23}, the black hole entropy \cite{R24}, the removal of the singularity of the early universe 
\cite{R25}, the Casimir effect in the rainbow Einstein's universe \cite{R8}, massive scalar field in RG Schwarzschild metric \cite{R26}, five-dimensional Yang--Mills black holes in massive RG \cite{R27}.

Moreover, recent studies are carried out on the effects of the RG on the dynamics of Klein-Gordon (KG) particles (i.e., spin-0 mesons), Dirac particles (spin-1/2 fermionic particles), and Duffen-Kemmer-Peatiau (DKP) particles (spin-1 particles like bosons and photons) in different spacetime backgrounds. For example, in a cosmic string spacetime background in rainbow gravity, Bezzerra et al. \cite{R81} have studied Landau levels via Schr\"{o}dinger and KG equations, Bakke and Mota \cite{R28} have studies the Dirac oscillator, they have also studied the Aharonov-Bohm effect \cite{R29}. Hosseinpour et al. \cite{R5} have studied the DKP-particles,, Sogut et al. \cite{R11} have studied the quantum dynamics of photon, and Kangal et al. \cite{R12} have studied KG-particles in a topologically trivial G\"{o}del-type spacetime in rainbow gravity. 

Very recently, position-dependent mass (PDM) concept  (e.g., \cite%
{R30,R31,R32,R33,R34,R35,R36,R37}) has been introduced to study PDM KG-oscillators in cosmic string spacetime within Kaluza-Klein theory \cite{R38}, in (2+1)-dimensional G\"{u}rses spacetime backgrounds \cite{R39}, in Minkowski spacetime with space-like dislocation \cite{R40}, and in PDM KG-Coulomb particles in cosmic string rainbow gravity spacetime and a uniform magnetic field \cite{R40.1}. In the later \cite{R40.1}, however, we have noticed that only one rainbow functions pair  (i.e., $g_{_{0}}\left( y\right) =1$, $g_{_{1}}\left( y\right) =\sqrt{1-\epsilon
y^{2}}$) provides invariance of the Planck's energy scale $E_p$ for both KG-particles and anti-particles. In the current proposal, nevertheless, we argue that as long as relativistic particles and anti-particles are in point, then $y=E/E_p$ should be fine tuned into $y=|E|/E_{p}$ so that   $0\leq \left( y=E/E_{p}\right) \leq 1$ is secured. Only under such fine tuning, the energies of the probe relativistic particles, $E=+|E|=E_+$ , and anti-particles, $E=-|E|=E_-$ , are equally treated within RG.  Through out the current methodical proposal, we use this fine tuning, therefore. 

Apriori, the cosmic string spacetime in rainbow gravity, using the natural units $%
c=\hbar =G=1$, takes the energy-dependent form%
\begin{equation}
ds^{2}=-\frac{1}{g_{_{0}}\left( y\right) ^{2}}dt^{2}+\frac{1}{g_{_{1}}\left(
y\right) ^{2}}\left( dr^{2}+\alpha ^{2}\,r^{2}d\varphi ^{2}+dz^{2}\right) ,
\label{e4}
\end{equation}%
where $\alpha=1-4G\mu $ is a constant related to the deficit angle of the conical spacetime,  $G$ is the Newton's constant, and $\mu $ is the linear mass density of the cosmic string so that $%
\alpha <1$.  The corresponding metric tensor $g_{\mu \nu }$ is given by%
\begin{equation}
g_{\mu \nu }=diag\left( -\frac{1}{g_{_{0}}\left( y\right) ^{2}},\frac{1}{%
g_{_{1}}\left( y\right) ^{2}},\frac{\alpha ^{2}\,r^{2}}{g_{_{1}}\left(
y\right) ^{2}},\frac{1}{g_{_{1}}\left( y\right) ^{2}}\right) ;\;\mu ,\nu
=t,r,\varphi ,z,  \label{e5}
\end{equation}%
with 
\begin{equation}
\det \left( g_{\mu \nu }\right) =-\frac{\alpha ^{2}\,r^{2}}{g_{_{0}}\left(
y\right) ^{2}g_{_{1}}\left( y\right) ^{6}}\Longrightarrow g^{\mu \nu
}=diag\left( -g_{_{0}}\left( y\right) ^{2},g_{_{1}}\left( y\right) ^{2},%
\frac{g_{_{1}}\left( y\right) ^{2}}{\alpha ^{2}\,r^{2}},g_{_{1}}\left(
y\right) ^{2}\right) .  \label{e6}
\end{equation}%
In the current methodical proposal, we study the effects of such cosmic string rainbow gravity on KG-particles in a non-uniform and a uniform magnetic field. In so doing, we shall be interested in three pairs of rainbow functions: (i) $g_{_{0}}\left( y\right) =1$, $g_{_{1}}\left( y\right) =\sqrt{1-\epsilon
y^{2}}$, and $g_{_{0}}\left( y\right) =1$, $g_{_{1}}\left( y\right) =\sqrt{%
1-\epsilon y}$, which belong to the set of rainbow functions $g_{_{0}}\left(
y\right) =1$, $g_{_{1}}\left( y\right) =\sqrt{1-\epsilon y^{n}}$ (where $%
\epsilon $ is the rainbow parameter and is a dimensionless constant of order unity) used to describe the geometry of spacetime in loop quantum gravity
\cite{R6,R61,R8,R41,R42,R421}, (ii) $%
g_{_{0}}\left( y\right) =g_{_{1}}\left( y\right) =\left( 1-\epsilon y\right)
^{-1}$, a suitable set used to resolve the horizon problem \cite{R13,R43}, and (iii) $g_{_{0}}\left( y\right) =\left( e^{\epsilon y}-1\right) /\epsilon
y$ and $g_{_{1}}\left( y\right) =1$, which are obtained from the spectra of
gamma-ray bursts at cosmological distances \cite{R6}. 

Our paper is organized as follows. In section 2, we discuss the KG-particles in the cosmic string rainbow gravity spacetime (\ref{e4}) in a non-uniform magnetic field  (i.e., $\mathbf{B}=\mathbf{\nabla }\times \mathbf{A}=\frac{3}{2}B_{\circ }r\,\hat{z}$ ). We bring the corresponding KG-equation into the one-dimensional form of the two-dimensional radial Schr\"{o}dinger oscillator equation. Hence the notion of KG-oscillators is unavoidable in the process. Using the above mentioned sets of the rainbow functions, we first discuss and report the effects of rainbow gravity on the energy levels of  the KG-oscillators. In section 3, we revisit and discuss (within the fine tuned $y$) the KG-Coulomb particles \cite{R40.1} in the cosmic string rainbow gravity spacetime (\ref{e4}) in a uniform magnetic field (i.e., $\mathbf{B}=\mathbf{\nabla }\times \mathbf{A}=\frac{1}{2}B_{\circ }\,\hat{z}$ ). In this case, KG-equation reduces into the one-dimensional form of the two-dimensional radial Schr\"{o}dinger Coulomb equation (hence the notion of KG-Coulombic particles is used). We again use the above mentioned pairs of rainbow function. We conclude in section 4.

\section{KG-oscillators in cosmic string rainbow gravity spacetime and a non-uniform magnetic field}

In the cosmic string rainbow gravity spacetime background (\ref{e4}), a KG-particle of charge $e$ in a 4-vector potential $A_{\mu }$ is described (in $c=\hbar =1$ units) by the KG-equation%
\begin{equation}
\frac{1}{\sqrt{-g}}D_{\mu }\left( \sqrt{-g}g^{\mu \nu }D_{\nu }\Psi \right)
=m^{2}\Psi ,  \label{e7}
\end{equation}%
where $D_{\mu }$ is the gauge-covariant derivative given by $D_{\mu
}=\partial _{\mu }-ieA_{\mu }$, and $m$ is the rest mass energy of the KG-particle.  Under such setting, our KG-equation (\ref{e7}) reads%
\begin{equation}
\left\{ -g_{_{0}}\left( y\right) ^{2}\partial _{t}^{2}+g_{_{1}}\left(
y\right) ^{2}\left[ \partial _{r}^{2}+\frac{1}{r}\partial _{r}+\frac{1}{\alpha ^{2}\,r^{2}}\left( \partial _{\varphi
}-ieA_{\varphi }\right) ^{2}+\partial _{z}^{2}\right] \right\} \Psi \left(
t,r,\varphi ,z\right) =m^{2}\Psi \left( t,r,\varphi ,z\right) ,  \label{e9}
\end{equation}%
We now use the substitution%
\begin{equation}
\Psi \left( t,r,\varphi ,z\right) =\exp \left( i\left[ \ell \varphi
+k_{z}z-Et\right] \right) \psi \left( r\right) ,  \label{e11}
\end{equation}%
in Eq. (\ref{e9}) to obtain%
\begin{equation}
\left\{ \tilde{E}^{2}+g_{_{1}}\left( y\right) ^{2}\left[ \partial _{r}^{2}+%
\frac{1}{r}\partial _{r} -\frac{\left( \ell -eA_{\varphi
}\right) ^{2}}{\alpha ^{2}\,r^{2}}\right] \right\} \psi \left( r\right) =0,
\label{e12}
\end{equation}%
where 
\begin{equation}
\tilde{E}^{2}=g_{_{0}}\left( y\right) ^{2}E^{2}-g_{_{1}}\left( y\right)
^{2}k_{z}^{2}-m^{2}  \label{e13}
\end{equation}

We now consider  $A_{\varphi }=\frac{1}{2}B_{\circ }r^{2}$ to yield a non-uniform magnetic field $\mathbf{B}=\mathbf{%
\nabla }\times \mathbf{A}=\frac{3}{2}B_{\circ }r\,\hat{z}$. Consequently, Eq.(\ref{e12}) becomes%
\begin{equation}
\left\{ \lambda +\partial _{r}^{2}+\frac{1}{r}\partial _{r}
-\frac{\tilde{\ell}^{2}}{r^{2}}-\frac{1}{4}\tilde{B}^{2}r^{2}\right\} \psi
\left( r\right) =0,  \label{e14}
\end{equation}%
where%
\begin{equation}
\lambda =\frac{g_{_{0}}\left( y\right) ^{2}E^{2}+g_{_{1}}\left( y\right)
^{2}\left( \tilde{B}\tilde{\ell}-k_{z}^{2}\right) -m^{2}}{g_{_{1}}\left(
y\right) ^{2}},\;\tilde{\ell}=\frac{\ell }{\alpha },\;\tilde{B}=\frac{%
eB_{\circ }}{\alpha }.  \label{e15}
\end{equation}%
Moreover, with $\psi \left( r\right) =R\left( r\right) /\sqrt{r}$ we obtain the two-dimensional radial KG-oscillators%
\begin{equation}
\left\{ \partial _{r}^{2}-\frac{\left( \tilde{\ell}^{2}-1/4\right) }{r^{2}}-%
\frac{1}{4}\tilde{B}^{2}r^{2}+\lambda \right\} R\left( r\right) =0.
\label{e17}
\end{equation}%
Which obviously admits exact solution in the form of hypergeometric function
so that%
\begin{equation}
R\left( r\right) =C\,r^{|\tilde{\ell}|+1/2}\exp \left( -\frac{|\tilde{B}|}{4}%
r^{2}\right) \,_{1}F_{1}\left( \frac{1}{2}+\frac{|\tilde{\ell}|}{2}-\frac{%
\lambda }{2|\tilde{B}|},1+|\tilde{\ell}|,\frac{|\tilde{B}|}{4}r^{2}\right) .
\label{e18}
\end{equation}%
However, to secure finiteness and square integrability we need to terminate the hypergeometric function into a polynomial of degree $n_{r}\geq 0$ so that the condition%
\begin{equation}
\frac{1}{2}+\frac{|\tilde{\ell}|}{2}-\frac{\lambda }{2|\tilde{B}|}=-n_{r}
\label{e19}
\end{equation}%
is satisfied. This would in turn imply that%
\begin{equation}
\lambda _{n_{r},\ell }=|\tilde{B}|\left( 2n_{r}+|\tilde{\ell}|+1\right)
,  \label{e20}
\end{equation}%
and%
\begin{equation}
 \psi(r)=\frac{R(r)}{\sqrt{r}} =C\,r^{|\tilde{\ell}|}\exp \left( -\frac{|\tilde{B}|%
}{4}r^{2}\right) \,_{1}F_{1}\left( -n_{r},1+|\tilde{\ell}|,\frac{|\tilde{B}|%
}{4}r^{2}\right) \label{e20.1} .
\end{equation}
Consequently, equation (\ref{e15}) would read%
\begin{equation}
g_{_{0}}\left( y\right) ^{2}E^{2}-m^{2}=g_{_{1}}\left( y\right)
^{2}\,K_{n_{r},\ell };\;K_{n_{r},\ell }=\left[ |\tilde{B}|\left( 2n_{r}+|%
\tilde{\ell}|+1\right) -\tilde{B}\tilde{\ell}+k_{z}^{2}\right] .  \label{e21}
\end{equation}
Before we proceed, it is convenient to observe that there are degeneracies associated with this relativistic energy relation (\ref{e21}). Obviously, all states with $\tilde{B}\tilde{\ell}=+|\tilde{B}||\tilde{\ell}|$ (i.e., for $\tilde{B}%
=+|\tilde{B}|=+ |e|B_{\circ }/\alpha$ and $\tilde{\ell}=+ |\tilde{%
\ell}|$ or for $\tilde{B}%
=- |\tilde{B}|=- |e|B_{\circ }/\alpha $ and $\tilde{\ell}=- |\tilde{%
\ell}|$) combine with $S$-state (i.e., $\ell =0;\,\tilde{\ell}=\ell
/\alpha .$), for a given radial quantum number $n_{r}$, so that the energy dispersion relation (\ref{e21}) reads%
\begin{equation}
g_{_{0}}\left( y\right) ^{2}E^{2}-m^{2}=g_{_{1}}\left( y\right) ^{2}\left[ |%
\tilde{B}|\left( 2n_{r}+1\right) +k_{z}^{2}\right] .  \label{e22}
\end{equation}%
Whereas, for $\tilde{B}\tilde{\ell}=-|\tilde{B}||\tilde{\ell}|$ (i.e., for $\tilde{B}%
=+|\tilde{B}|=+ |e|B_{\circ }/\alpha$ and $\tilde{\ell}=-|\tilde{%
\ell}|$ or for $\tilde{B}%
=- |\tilde{B}|=- |e|B_{\circ }/\alpha $ and $\tilde{\ell}=+|\tilde{\ell}|$) the energy dispersion relation (\ref{e21}) yields%
\begin{equation}
g_{_{0}}\left( y\right) ^{2}E^{2}-m^{2}=g_{_{1}}\left( y\right) ^{2}\left[ |%
\tilde{B}|\left( 2n_{r}+2|\tilde{\ell}|+1\right) +k_{z}^{2}\right] .
\label{e23}
\end{equation}%
Which suggests that for $\ell \neq 0$ there are degeneracies for every magnetic quantum number
$\ell=\pm 1,\pm 2,\cdots$. Such degeneracies may very well be called charge associated degeneracies. At this point, it should be made clear that such degeneracies have nothings to do with the rainbow gravity effects as can be concluded from (\ref{e21}). In what follows, however, we shall only consider positively charged KG-oscillators
so that (\ref{e21}) yields%
\begin{equation}
g_{_{0}}\left( y\right) ^{2}E^{2}-m^{2}=g_{_{1}}\left( y\right) ^{2}\left[ |%
\tilde{B}|\left( 2n_{r}+1\right) +k_{z}^{2}\right]   \label{e221}
\end{equation}%
for $\ell =+|\ell |$, and%
\begin{equation}
g_{_{0}}\left( y\right) ^{2}E^{2}-m^{2}=g_{_{1}}\left( y\right) ^{2}\left[ |%
\tilde{B}|\left( 2n_{r}+2|\tilde{\ell}|+1\right) +k_{z}^{2}\right] 
\label{e231}
\end{equation}
for $\ell =-|\ell |$. Consequently, equation (\ref{e221}) would allow all positive/negative energies with $\ell =+|\ell |$ to combine with the corresponding positive/negative $S$-states (i.e., $\ell=0$ states) for a given $n_r$. Whereas, equation (\ref{e231}) would allow the  positive/negative energies with $\ell =-|\ell |$ to appear in the corresponding energy spectra.

We may at this point consider different rainbow functions and discuss their effects on the energy levels of 
(\ref{e21}).%

\subsection{The rainbow functions   $g_{_{0}}\left( y\right) =1$ and $g_{_{1}}\left( y\right) =%
\sqrt{1-\epsilon y^{n}}$}
This set of rainbow functions are motivated from loop quantum gravity \cite{R6,R61,R8,R9}. In the current study, however, we shall consider $n=1$ and $n=2$ that are commonly used in similar studies 
(c.f., e.g., \cite{R8,R46}). 

\subsubsection{$n=2$ case:}

We first start with the rainbow functions pair   $g_{_{0}}\left( y\right) =1$ and $g_{_{1}}\left( y\right) =%
\sqrt{1-\epsilon y^{2}}$ (i.e., for $n=2$).
Such rainbow functions pair would, using (\ref{e21}), result%
\begin{equation}
E^{2}-m^{2}=\left( 1-\epsilon \frac{E^{2}}{E_{p}^{2}}\right) K_{n_{r},\ell
}\Longrightarrow E_\pm=\pm \sqrt{\frac{K_{n_{r},\ell }+m^{2}}{1+\delta
\,K_{n_{r},\ell }}};\;\delta =\frac{\epsilon }{E_{p}^{2}}.  \label{e24}
\end{equation}%
One should notice that an expansion about $\delta=0$ would imply 
\begin{equation}
   |E_\pm| \approx   \sqrt{K_{n_{r},\ell }+m^{2}}-\frac{1}{2}\sqrt{K_{n_{r},\ell }+m^{2}}K_{n_{r},\ell }\delta +O(\delta^2)<\sqrt{K_{n_{r},\ell }+m^{2}}, \label{e24.02}
\end{equation}
where $\sqrt{K_{n_{r},\ell }+m^2}$ is the exact energy without rainbow gravity. Hence, the energies of the probe KG-oscillators is less than the case  without rainbow gravity.

\begin{figure}[!ht]  
\centering
\includegraphics[width=0.3\textwidth]{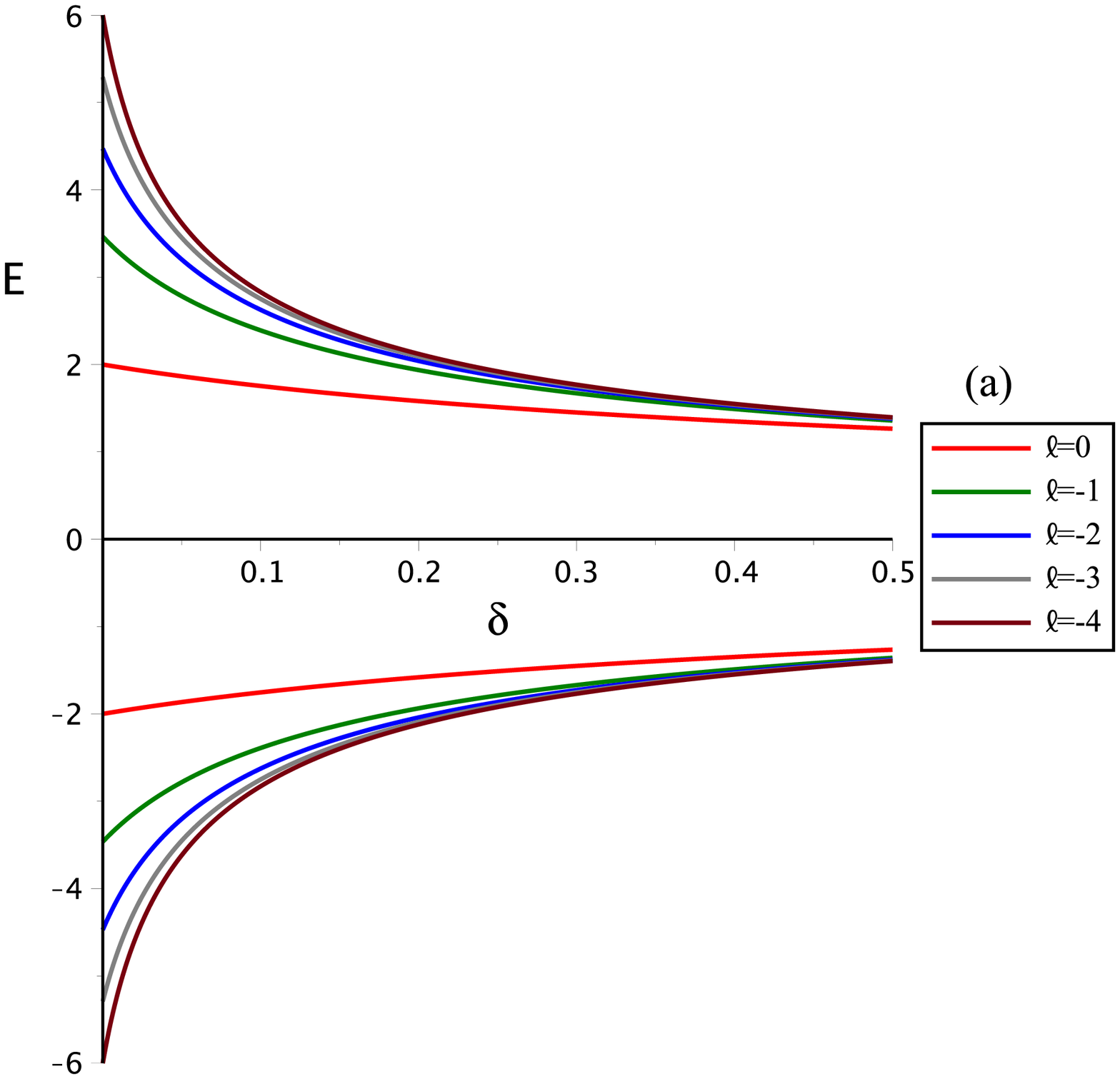}
\includegraphics[width=0.3\textwidth]{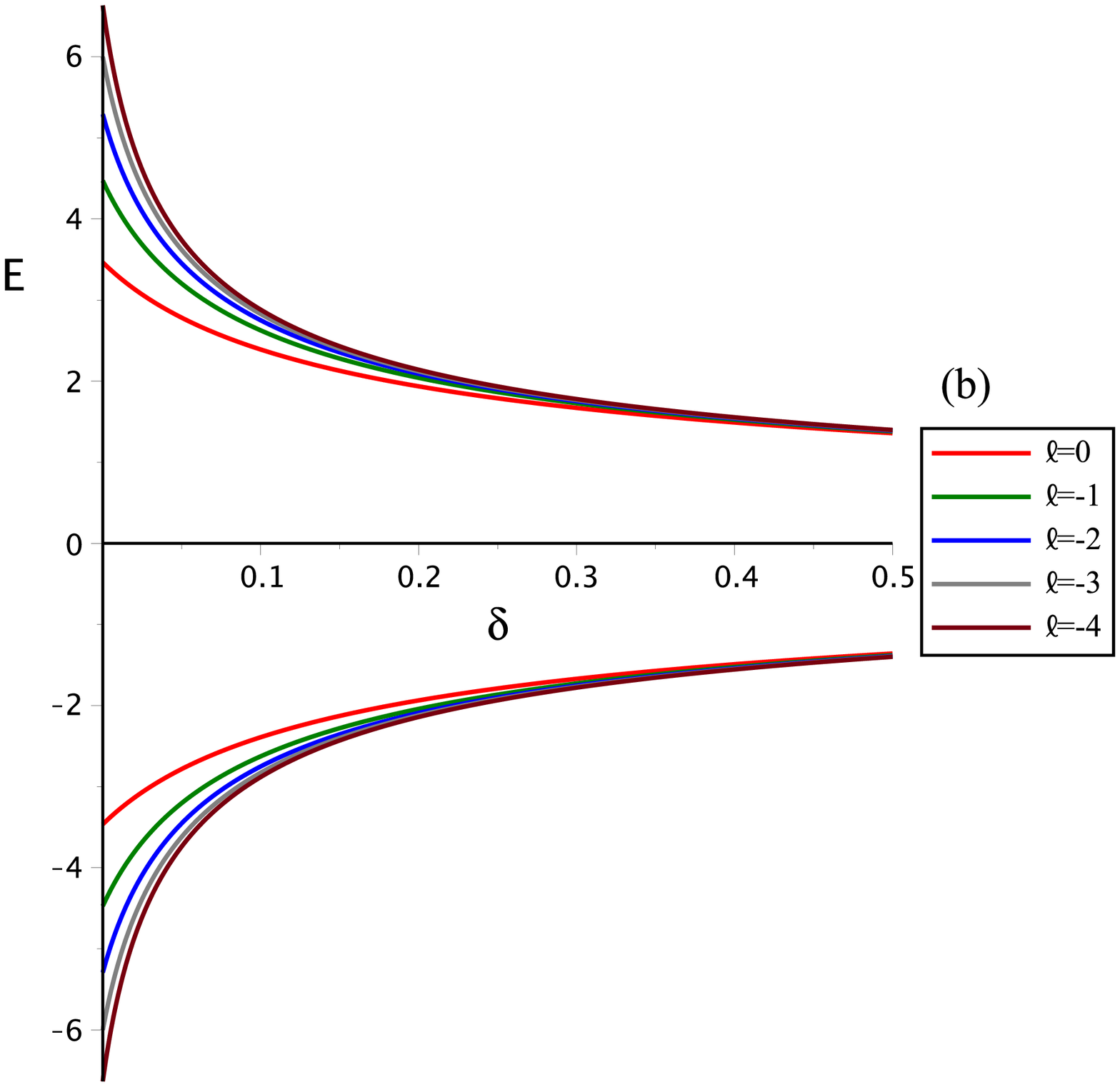} 
\includegraphics[width=0.3\textwidth]{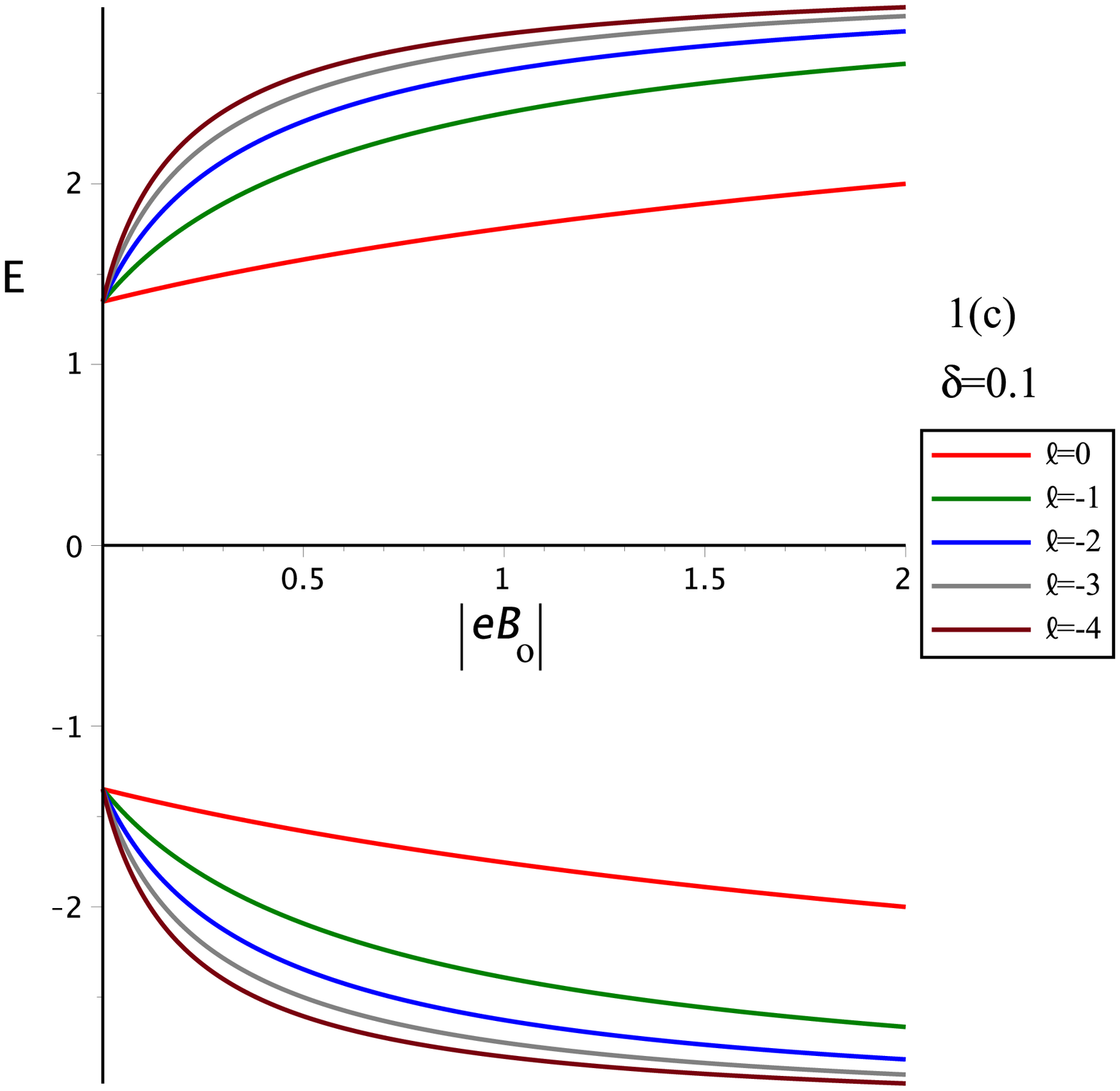}
\caption{\small 
{ The energy levels of (\ref{e24}), using $\alpha =1/2$, $%
m=k_{z}=1$, so that (a) shows $E$ against $\delta =\epsilon /E_{p}^{2}$ for $%
|eB_{\circ }|=1$, $n_{r}=0$, $\ell =0,-1,-2,-3,-4$, (b) shows $E$ against $%
\delta =\epsilon /E_{p}^{2}$ for $|eB_{\circ }|=1$, $n_{r}=2$, $\ell
=0,-1,-2,-3,-4$, and (c) shows $E$ against $|eB_{\circ }|$ for $\delta =0.1$%
, $n_{r}=0$, $\ell =0,-1,-2,-3,-4$.}}
\label{fig1}
\end{figure}%

We plot the corresponding energies against $\delta
=\epsilon /E_{p}^{2}$ in Figures 1(a), and (b). We observe that, for a given radial quantum number $n_{r}$, eminent clustering of positive/negative energy levels as $\delta $ grows up from zero.  In Figure 1(c), moreover, we plot the energies against $\left\vert eB_{\circ }\right\vert $. It is obvious that as $\left\vert eB_{\circ }\right\vert \rightarrow 0$ the energy levels converge to the values $E_\pm\sim \pm \sqrt{\left(
k_{z}^{2}+m^{2}\right) /\left( 1+\delta k_{z}^{2}\right) }=\pm \sqrt{2/1.1}\sim \pm1.35$ (for $\delta =0.1$, and $m=k_{z}=1$ value used here). That is, at this limit positive/negative energy states emerge from the same positive/negative values at $B_{\circ }=0$  irrespective of the values of the radial and magnetic quantum numbers $n_{r}$ and $\ell $, respectively. On the other hand, as $\left\vert eB_{\circ }\right\vert \rightarrow \infty$, the energy levels cluster about $E_\pm\sim \pm \sqrt{1/\delta }$ . That is, 
\begin{equation}
   \lim\limits_{|\tilde{B}|\rightarrow \infty }E_{\pm}\approx \pm\frac{1}{\sqrt{\delta}}=\pm \frac{E_p}{\sqrt{\epsilon}}. \label{e24.01}
\end{equation}
This is documented in Figure 1(c)  as the energies tend to approach the value 
$|E_{n_{r},\ell }|\sim \sqrt{1/\delta }= \sqrt{10}\sim 3.16$ for $\left\vert eB_{\circ }\right\vert >>1$ and $\delta =0.1$. Interestingly, we observe that under such rainbow functions structure the energy levels are destined to be within the range
\begin{equation}
 \sqrt{( k_{z}^{2}+m^{2}) /( 1+\delta
k_{z}^{2}) }\leq |E_\pm|\,\leq \sqrt{1/\delta }=E_{p}/\sqrt{%
\epsilon }.   \label{e24.1}
\end{equation}
This relation suggests that the rainbow parameter $\epsilon\geq 1$. However, it also mandates an upper limit $|E_\pm|_{max}\,= E_{p}$ \cite{R44}  for the energies of the probe KG-oscillators (for the rainbow parameter  $\epsilon$ of order one). This is consistent with the DSR or the rainbow model. Yet, to switch off rainbow gravity and return back to cosmic string spacetime in GR, we set $\epsilon=0 \rightarrow \delta=0$. 

\subsubsection{$n=1$ case:}

Next we consider the rainbow functions pair $g_{_{0}}\left( y\right) =1$ and $g_{_{1}}\left( y\right) =\sqrt{%
1-\epsilon y}$ (i.e., $n=1$). 
Such rainbow functions model in (\ref{e21}) would imply%
\begin{equation}
E_{\pm}^{2}-m^{2}=\left( 1-\epsilon \frac{|E|}{E_{p}}\right) K_{n_{r},\ell
}\Longrightarrow E_{\pm}=\mp\beta K_{n_{r},\ell }\pm \sqrt{\beta ^{2}K_{n_{r},\ell
}^{2}+K_{n_{r},\ell }+m^{2}};\;\beta =\frac{\epsilon }{2E_{p}},  \label{e25}
\end{equation}%
\begin{figure}[!ht]  
\centering
\includegraphics[width=0.3\textwidth]{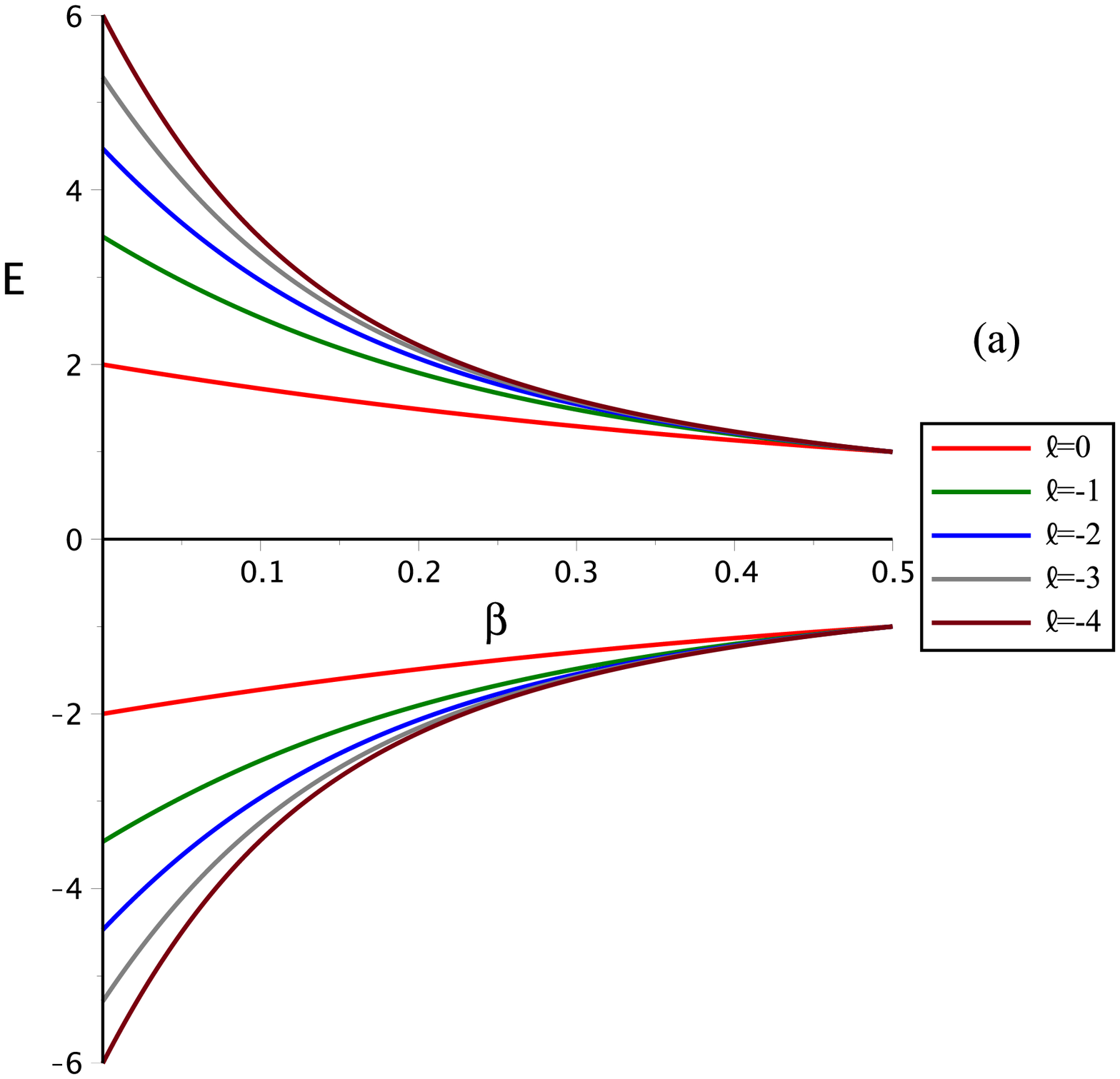}
\includegraphics[width=0.3\textwidth]{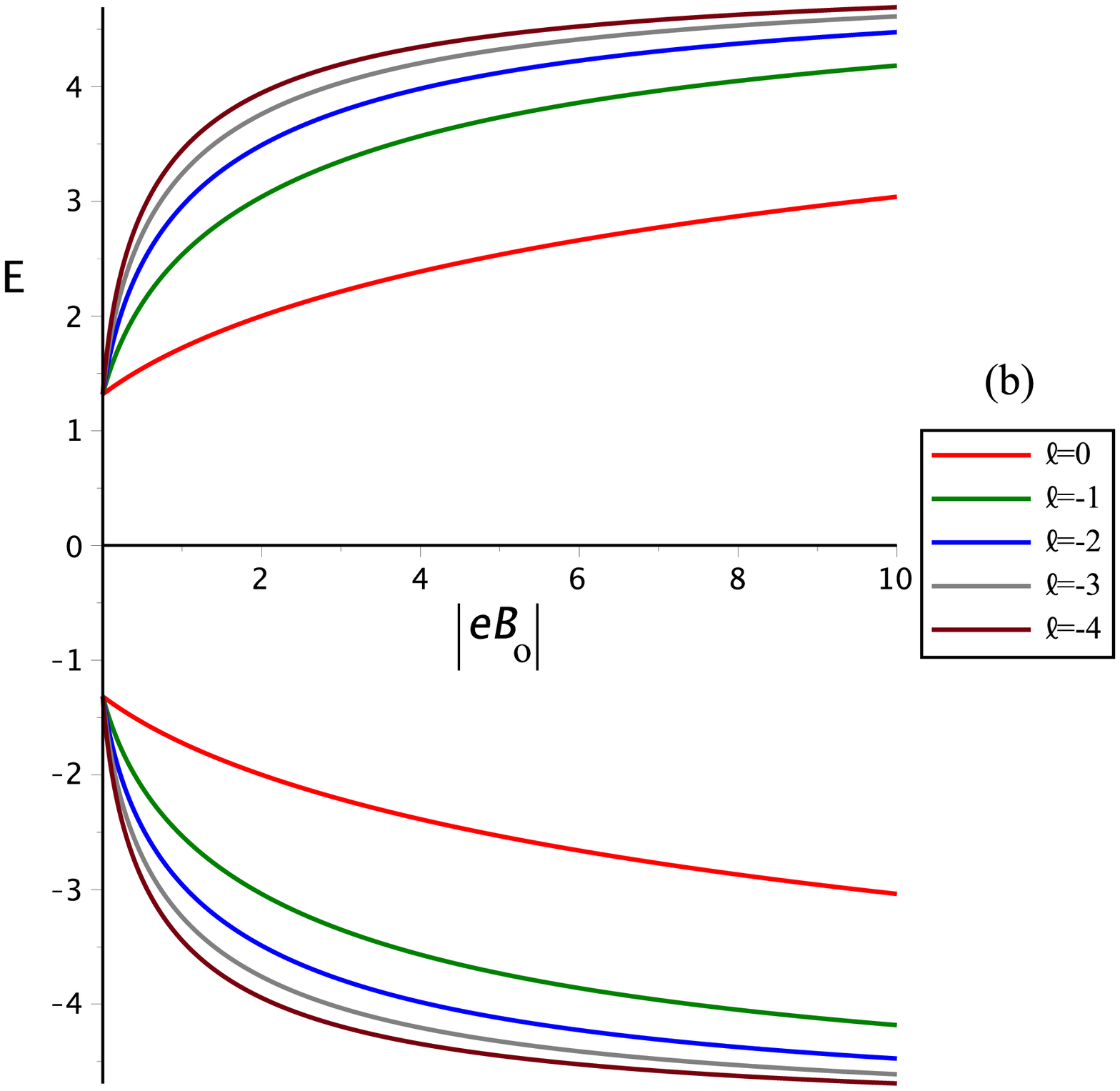} 
\caption{\small 
{ The energy levels of (\ref{e25}), using $\alpha =1/2$, $%
m=k_{z}=1$, so that (a) shows $E$ against $\beta =\epsilon /2E_{p}$ for $%
|eB_{\circ }|=1$, $n_{r}=0$, $\ell =0,-1,-2,-3,-4$, and (b) shows $E$
against $|eB_{\circ }|$ for $\beta =0.1$, $n_{r}=0$, $\ell =0,-1,-2,-3,-4$.}}
\label{fig2}
\end{figure}%
where, we have used $|E|=\pm E_\pm$. Moreover,  one may expand (\ref{e25}) about $\beta=0$ to obtain
\begin{equation}
   |E_{\pm}|\simeq \sqrt{K_{n_{r},\ell }+m^2}-K_{n_{r},\ell }\beta +O(\beta^2)<\sqrt{K_{n_{r},\ell }+m^2}, \label{e25.2}
\end{equation}
where $\sqrt{K_{n_{r},\ell }+m^2}$ is the exact energy in no rainbow gravity.%

In Figures 2(a) and (b), we plot the energy levels against $\beta =\epsilon
/2E_{p}$ and $\left\vert eB_{\circ }\right\vert $, respectively. It is obvious that the energy levels are symmetric  about $E=0$. Yet, in Fig. 2(b) we observe that the asymptotic tendency of the energies as $\left\vert eB_{\circ }\right\vert\rightarrow \infty $ is%
\begin{equation}
    \lim\limits_{|\tilde{B}|\rightarrow \infty }E_{\pm}\approx \pm\frac{1}{2\beta}=\pm\frac{E_p}{\epsilon }. \label{e25.1}
\end{equation}
This would suggest that $\lim\limits_{|\tilde{B}| \rightarrow \infty }E_{\pm}\approx \pm 1/2\beta=\pm5$ for $\beta=0.1$ and hence  $|E_{\pm}|_{max}=1/2\beta=E_p$ for $\epsilon=1$  (i.e., consistent with DSR/rainbow gravity model \cite{R44} provided that $\epsilon \geq1$).  

\subsection{The of rainbow functions $g_{_{0}}\left( y\right) =g_{_{1}}\left( y\right) =\left(
1-\epsilon y\right) ^{-1}$}

Such rainbow functions assumption in (\ref{e21}) yields%
\begin{equation}
E^{2}-K_{n_{r},\ell }=\left( 1-\epsilon \frac{|E|}{E_{p}}\right)
^{2}m^{2}\Longrightarrow E_\pm=\frac{\mp m\gamma \pm \sqrt{K_{n_{r},\ell }\left(
1-\gamma ^{2}\right) +m^{2}}}{1-\gamma ^{2}};\;\gamma =\frac{\epsilon m}{%
E_{p}}<1.  \label{e26}
\end{equation}%

In Figures 3(a) we plot the energy levels against $\gamma =\epsilon m/E_{p}<1
$ to observe the rainbow gravity effect. In Figures 3(b), for $\gamma=0.1$, and 3(c), for $\gamma=0.8$, the energy levels are plotted against $\left\vert eB_{\circ
}\right\vert $ so that the magnetic field effect on the energy levels is shown. The energy levels are observed to preserve their symmetry about $E=0$ value.  Yet, we observe that as $\gamma$ increases from zero, the energy gap narrows down (i.e., for $\gamma=0.1$ in 3(b) the energy gap at $|eB_\circ|=0$ is $\approx 0.909$ whereas for $\gamma=0.8$ the energy gap at $|eB_\circ|=0$ is $\approx 0.556$ as documented in Fig. 3(b) and 3(c), respectively). Although such a rainbow function pair yields
\begin{equation}
  \lim\limits_{{\gamma}\rightarrow 0 }|E_{\pm}|\approx \sqrt{K_{n_{r},\ell }+m^{2}} -m\gamma+O(\gamma^2)<|E_\pm|_{\epsilon=0}=\sqrt{K_{n_{r},\ell }+m^{2}}, \label{e26.1}
\end{equation}%

\begin{figure}[!ht]  
\centering
\includegraphics[width=0.3\textwidth]{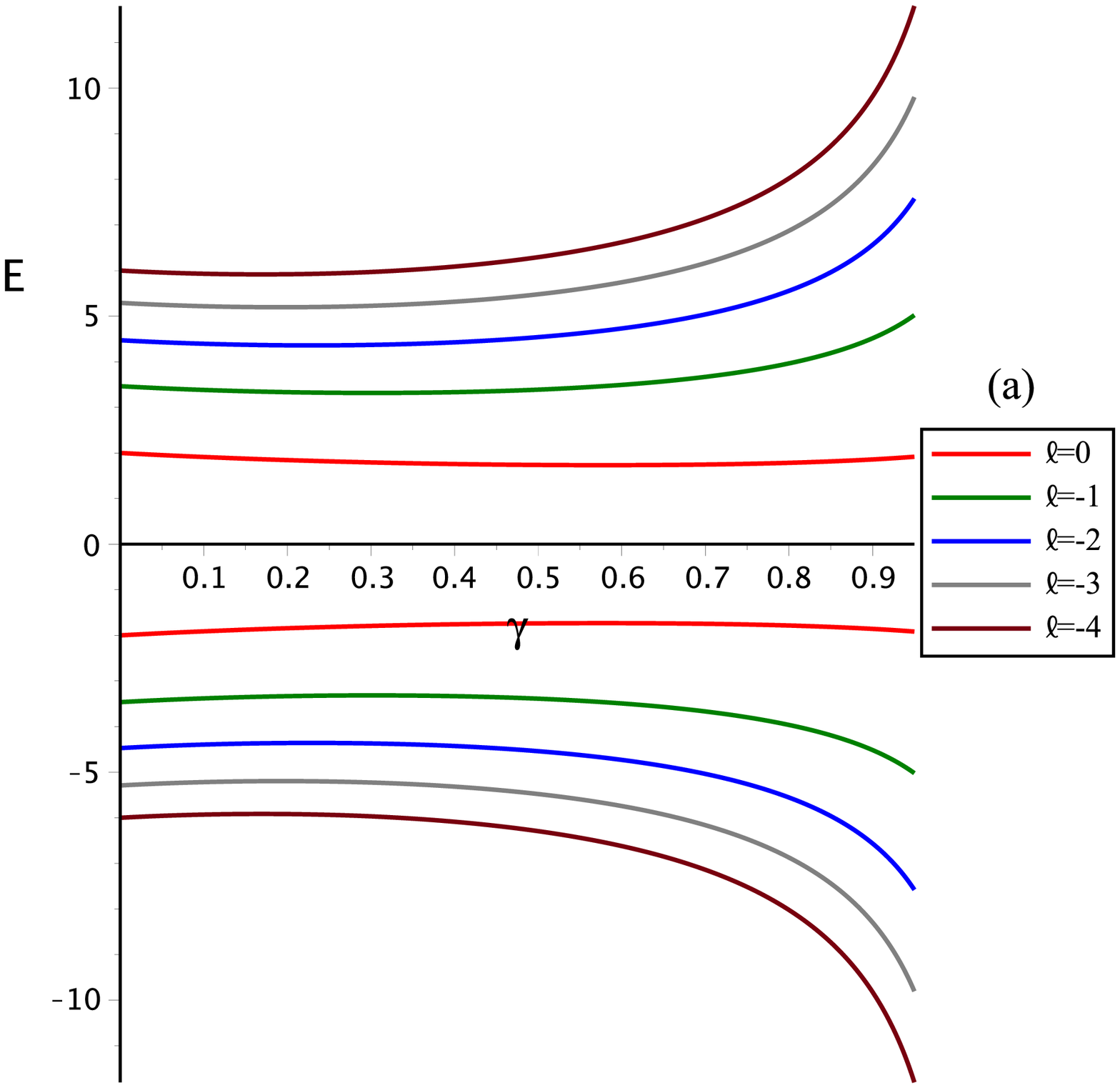}
\includegraphics[width=0.3\textwidth]{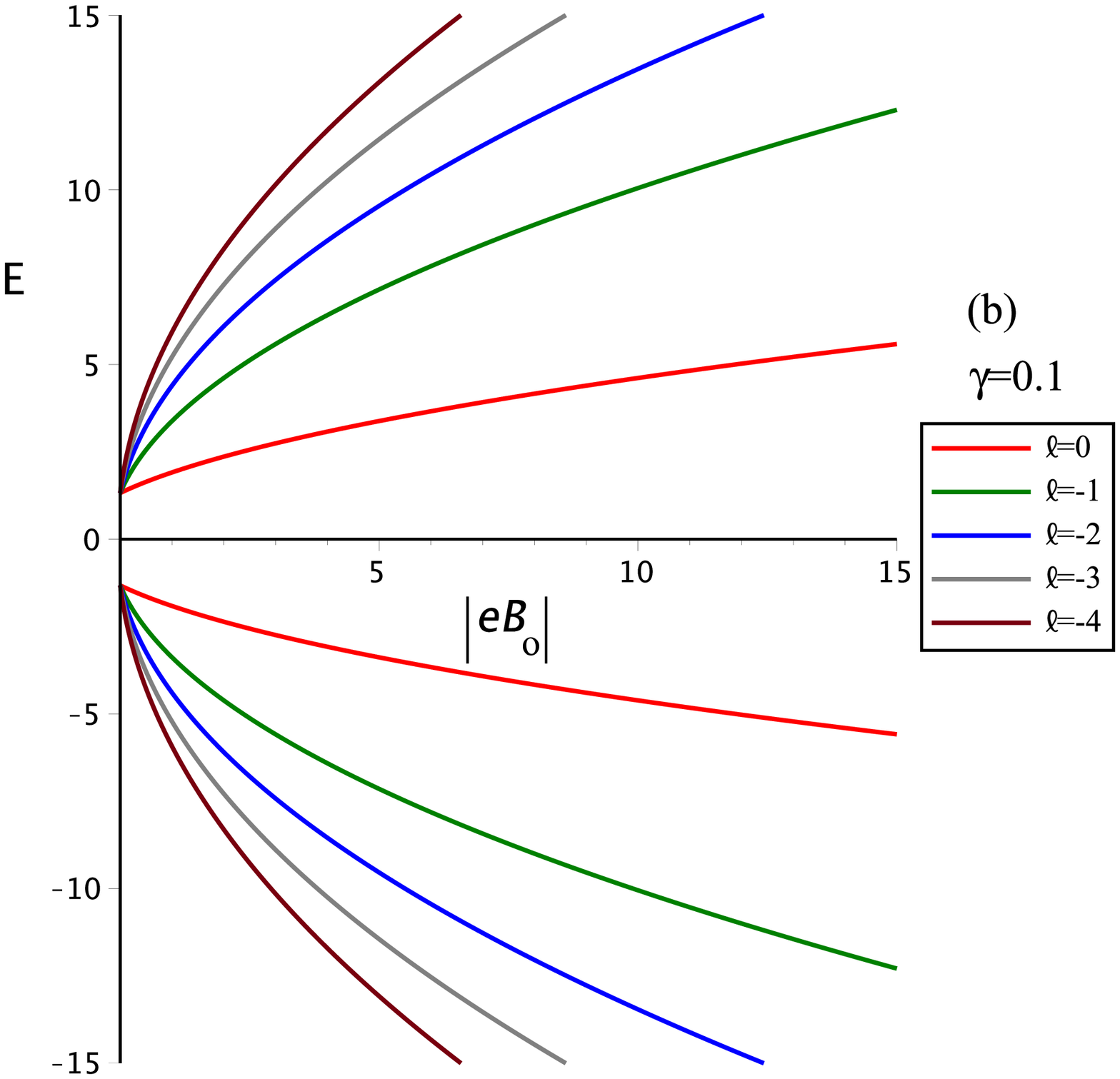} 
\includegraphics[width=0.3\textwidth]{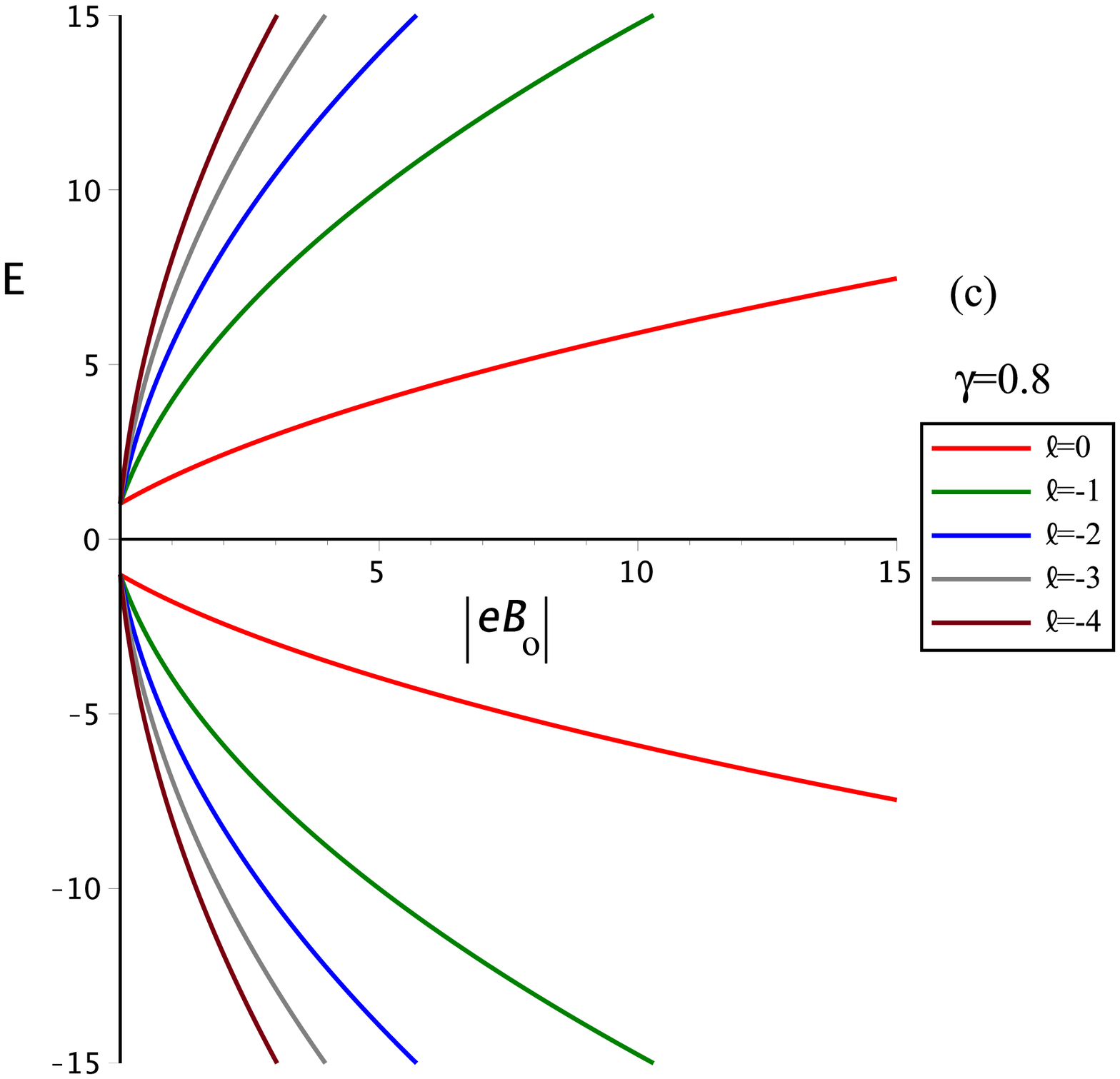}
\caption{\small 
{ The energy levels of (\ref{e26}), using $\alpha =1/2$, $%
m=k_{z}=1$, so that (a) shows $E$ against $\gamma =\epsilon m/E_{p}<1$ for $%
|eB_{\circ }|=1$, $n_{r}=0$, $\ell =0,-1,-2,-3,-4$, (b) shows $E$ against $%
|eB_{\circ }|$ for $\gamma =0.1$, $n_{r}=0$, $\ell =0,-1,-2,-3,-4$, and (c)
shows $E$ against $|eB_{\circ }|$ for $\gamma =0.8$, $n_{r}=0$, $\ell
=0,-1,-2,-3,-4$.}}
\label{fig3}
\end{figure}%
\begin{figure}[!ht]  
\centering
\includegraphics[width=0.3\textwidth]{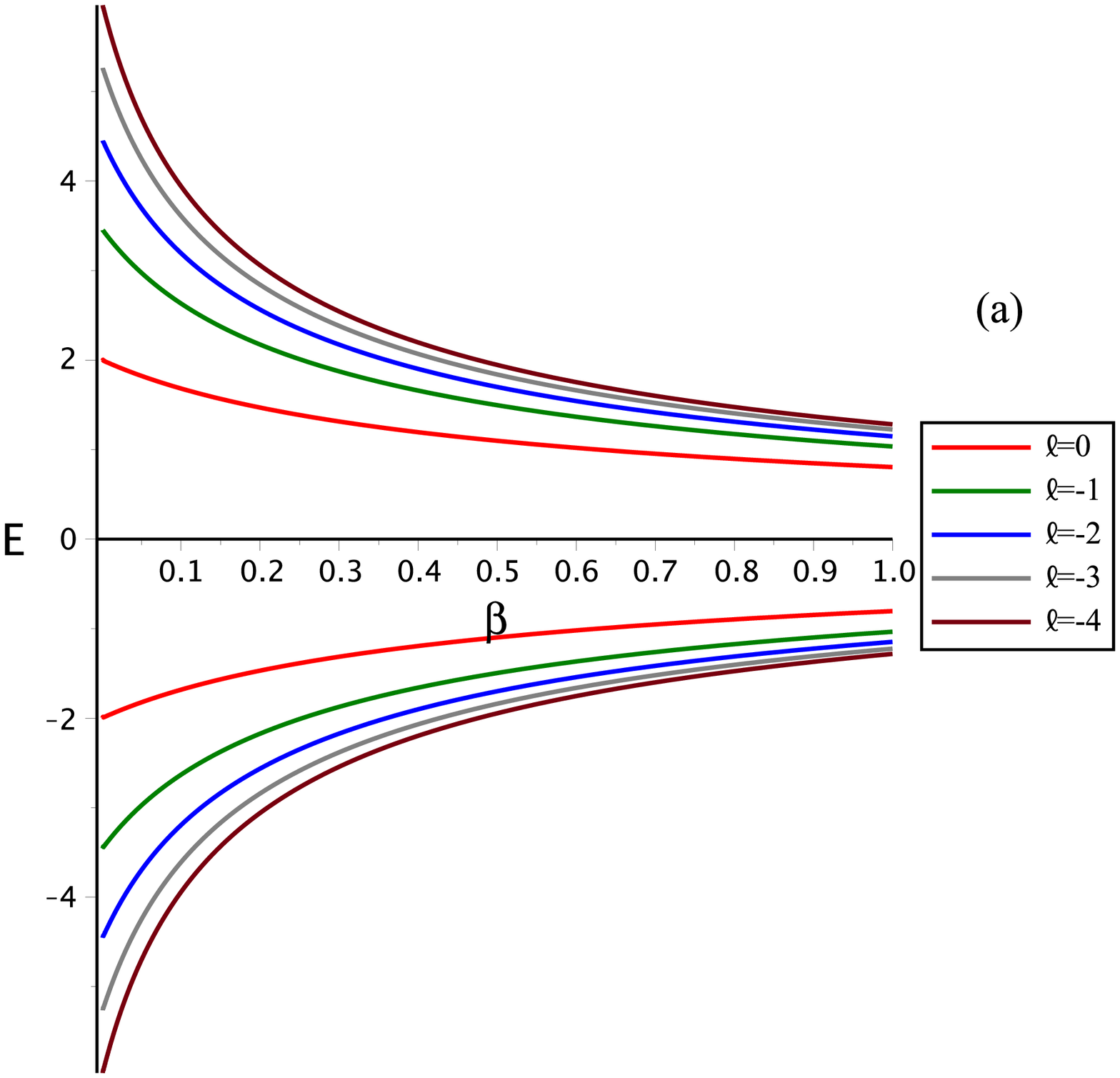}
\includegraphics[width=0.3\textwidth]{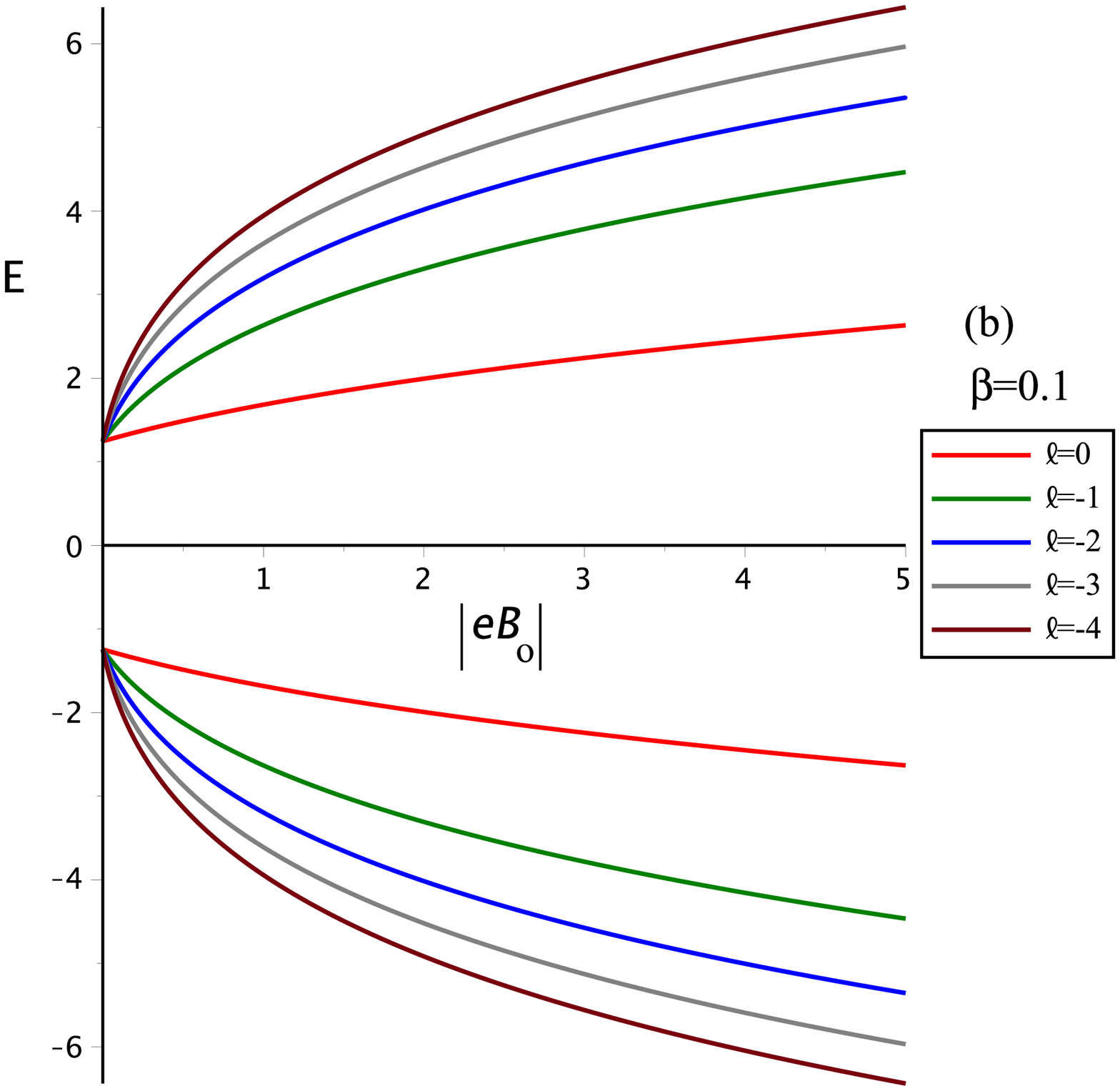}
\includegraphics[width=0.3\textwidth]{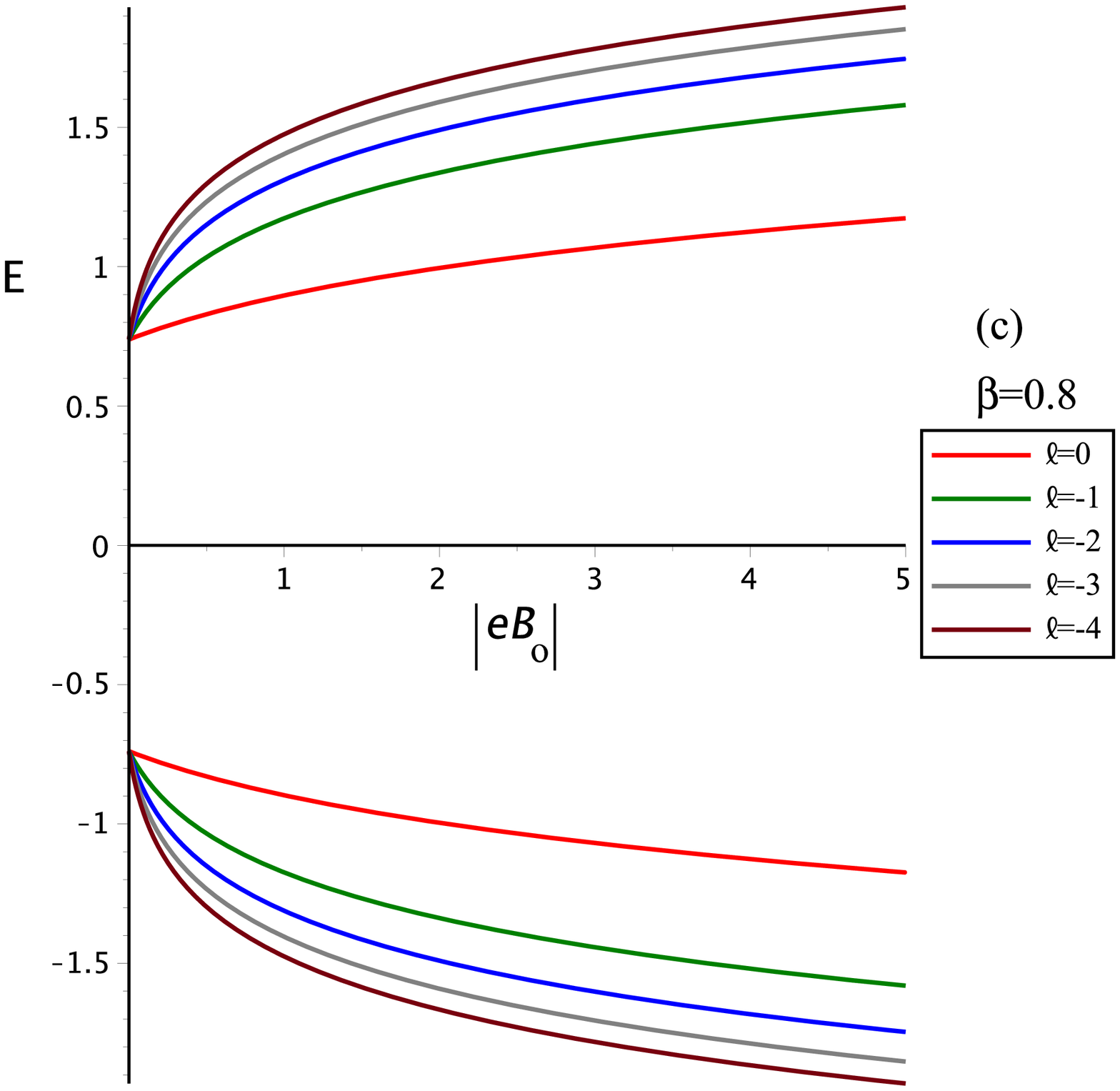}
\caption{\small 
{ The energy levels of (\ref{e27}), using $\alpha =1/2$, $%
m=k_{z}=1$, so that (a) shows $E$ against $\beta =\epsilon /2E_{p}$ for $%
|eB_{\circ }|=1$, $n_{r}=0$, $\ell =0,-1,-2,-3,-4$,  (b) shows $E$
against $|eB_{\circ }|$ for $\beta =0.1$, $n_{r}=0$, $\ell =0,-1,-2,-3,-4$, and (c) shows $E$
against $|eB_{\circ }|$ for $\beta =0.8$, $n_{r}=0$, $\ell =0,-1,-2,-3,-4$,. }}
\label{fig4}
\end{figure}%
it fails to show any eminent convergence of the energies towards the Planck's energy scale $E_p$. We may also observe that the tendency of the energy states to fly away to $\pm\infty$ and disappear from the spectrum is attributed to the singularity at $\gamma=1$ of the energies in (\ref{e26}). %

\subsection{The rainbow functions $g_{_{0}}\left( y\right) =\left( e^{\epsilon y}-1\right)
/\epsilon y$, and $g_{_{1}}\left( y\right) =1$}

Using such a rainbow functions structure in (\ref{e21}) yields%
\begin{equation}
E^{2}\left( \frac{e^{\epsilon |E|/E_{p}}-1}{\epsilon |E|/E_{p}}\right)
^{2}-m^{2}=K_{n_{r},\ell }\Longrightarrow E_\pm=\pm\frac{1}{2\beta }\ln \left( 1+ 
2\beta \sqrt{\left( K_{n_{r},\ell }+m^{2}\right) }\right) ;\;\beta =%
\frac{\epsilon }{2E_{p}}  \label{e27}
\end{equation}%
In Figure 4(a) we plot the energy levels against $\beta =\epsilon /2E_{p}$ . In Figures 4(b) (for $\beta=0.1$) and 4(c) (for $\beta=0.8$), we show the energy levels against $\left\vert eB_{\circ}\right\vert$. We observe that the energies are symmetric about $E=0$ value and an expansion about $\beta\rightarrow0$ implies
\begin{equation}
  |E_{\pm}|\approx \sqrt{K_{n_{r},\ell }+m^{2}} -m\beta+O(\beta^2)<|E_\pm|_{\epsilon=0}=\sqrt{K_{n_{r},\ell }+m^{2}}. \label{e27.1}
\end{equation}
Moreover,  it is clear that a comparison between 4(b) and 4(c) suggests that the energy gap narrows down 
as $\beta$ increases from zero.  Obviously, this rainbow functions pair does not show any eminent tendency towards the Planck's energy scale $E_p$.

\section{KG-Coulomb particles in cosmic string rainbow gravity spacetime in a uniform magnetic field; revisited}

In a recent paper \cite{R40.1}, we have studied PDM KG-Coulomb particles in cosmic string rainbow gravity and a uniform magnetic field, where $\mathbf{B}=\mathbf{\nabla }\times \mathbf{A}=\frac{1}{2}B_{\circ }\,\hat{z}$ is introduced by the electromagnetic vector potential $A_{\varphi }=\frac{1}{2}B_{\circ }r$. Therein, we have found that only one rainbow functions pair (i.e., $g_{_{0}}\left( y\right) =1$, $%
g_{_{1}}\left( y\right) =\sqrt{1-\epsilon y^{2}}$ ) complies with the the Planck's energy scale $E_p$ invariance (this is attributed to $y^2$ structure of said the rainbow function).  Therefore, this section is intended to show that the current fine tuning, $y=E/E_p\rightarrow y=|E|/E_p$, would indeed yield energy levels that are bounded between $|E_\pm| \leq |E_p|$. 

A substitution of vector potential $A_{\varphi }=\frac{1}{2}B_{\circ }r$ in (\ref{e12})  would immediately introduce a two-dimensional Schr\"{o}dinger-Coulomb like equation%
\begin{equation}
\left\{ \partial _{r}^{2}-\frac{\left( \tilde{\ell}^{2}-1/4\right) }{r^{2}}+%
\frac{\tilde{\ell}\,\tilde{B}}{r}+\Lambda \right\} R\left( r\right) =0,
\label{e28}
\end{equation}%
Where,%
\begin{equation}
\Lambda =\frac{g_{_{0}}\left( y\right) ^{2}E^{2}-g_{_{1}}\left( y\right)
^{2}\left( k_{z}^{2}+\frac{\tilde{B}^{2}}{4}\right) -m^{2}}{g_{_{1}}\left(
y\right) ^{2}},\;\tilde{\ell}=\frac{\ell }{\alpha },\;\tilde{B}=\frac{%
eB_{\circ }}{\alpha }.  \label{e28.1}
\end{equation}%
Hence the notion KG-Coulombic particles is unavoidable in the process. Equation (\ref{e28}) has an exact textbook solution in the form of hypergeometric functions%
\begin{equation}
R\left( r\right) \sim \,\left( 2i\sqrt{\Lambda }r\right) ^{|\tilde{\ell}%
|+1/2}\exp \left( -i\sqrt{\Lambda }r\right) \,_{1}F_{1}\left( \frac{1}{2}+%
|\tilde{\ell}|-\frac{\tilde{\ell}\,\tilde{B}}{2i\sqrt{\Lambda }},1+2|\tilde{%
\ell}|,2i\sqrt{\Lambda }r\right) .  \label{e29}
\end{equation}%
Nevertheless, the finiteness and square integrability of the $R(r)$ mandates that the hypergeometric series should be truncated into a polynomial of degree $n_{r}\geq 0$ so that the condition $1/2+|\tilde{\ell}|-\tilde{\ell}\,\tilde{B}/(2i\sqrt{\Lambda })%
=-n_{r}$ is satisfied. Then we obtain,%
\begin{equation}
i\sqrt{\Lambda }=\frac{\tilde{\ell}\,\tilde{B}}{2\tilde{n}};\;\tilde{n}%
=n_{r}+|\tilde{\ell}|+\frac{1}{2}\Rightarrow \Lambda _{n_{r},\ell }=-\frac{%
\tilde{\ell}\,^{2}\tilde{B}^{2}}{4\tilde{n}^{2}},  \label{e30}
\end{equation}%
and%
\begin{equation}
\psi \left( r\right) =\frac{R\left( r\right) }{\sqrt{r}}=\mathcal{N}\,r^{|%
\tilde{\ell}|}\exp \left( -\frac{|\tilde{\ell}\,\tilde{B}|}{2\tilde{n}}%
\,r\right) \,_{1}F_{1}\left( -n_{r},1+2|\tilde{\ell}|,\frac{|\tilde{\ell}\,%
\tilde{B}|}{\tilde{n}}r\right) .  \label{e31}
\end{equation}%
Consequently, Eq.(\ref{e28.1}) would read%
\begin{figure}[!ht]  
\centering
\includegraphics[width=0.35\textwidth]{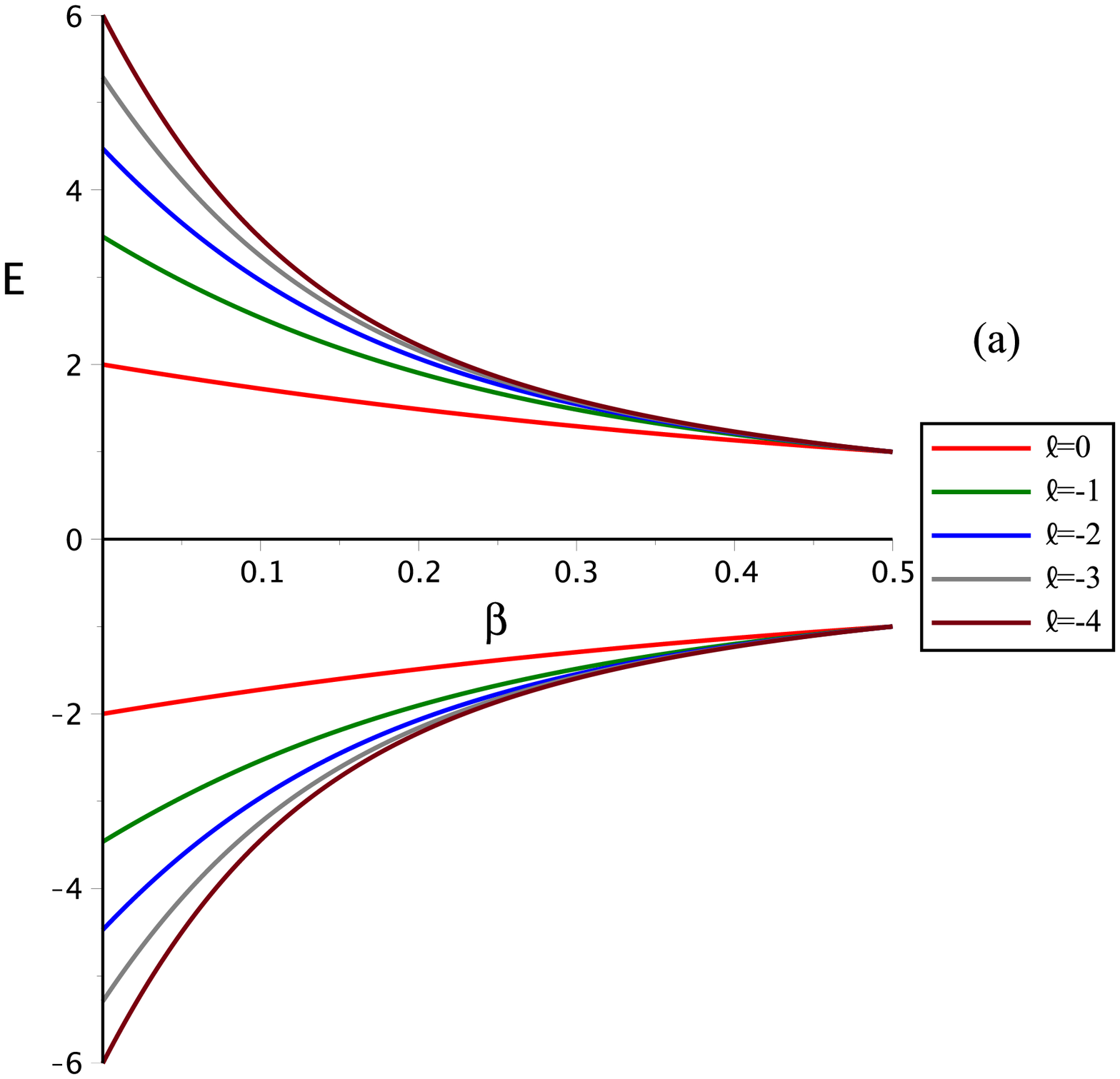}
\includegraphics[width=0.35\textwidth]{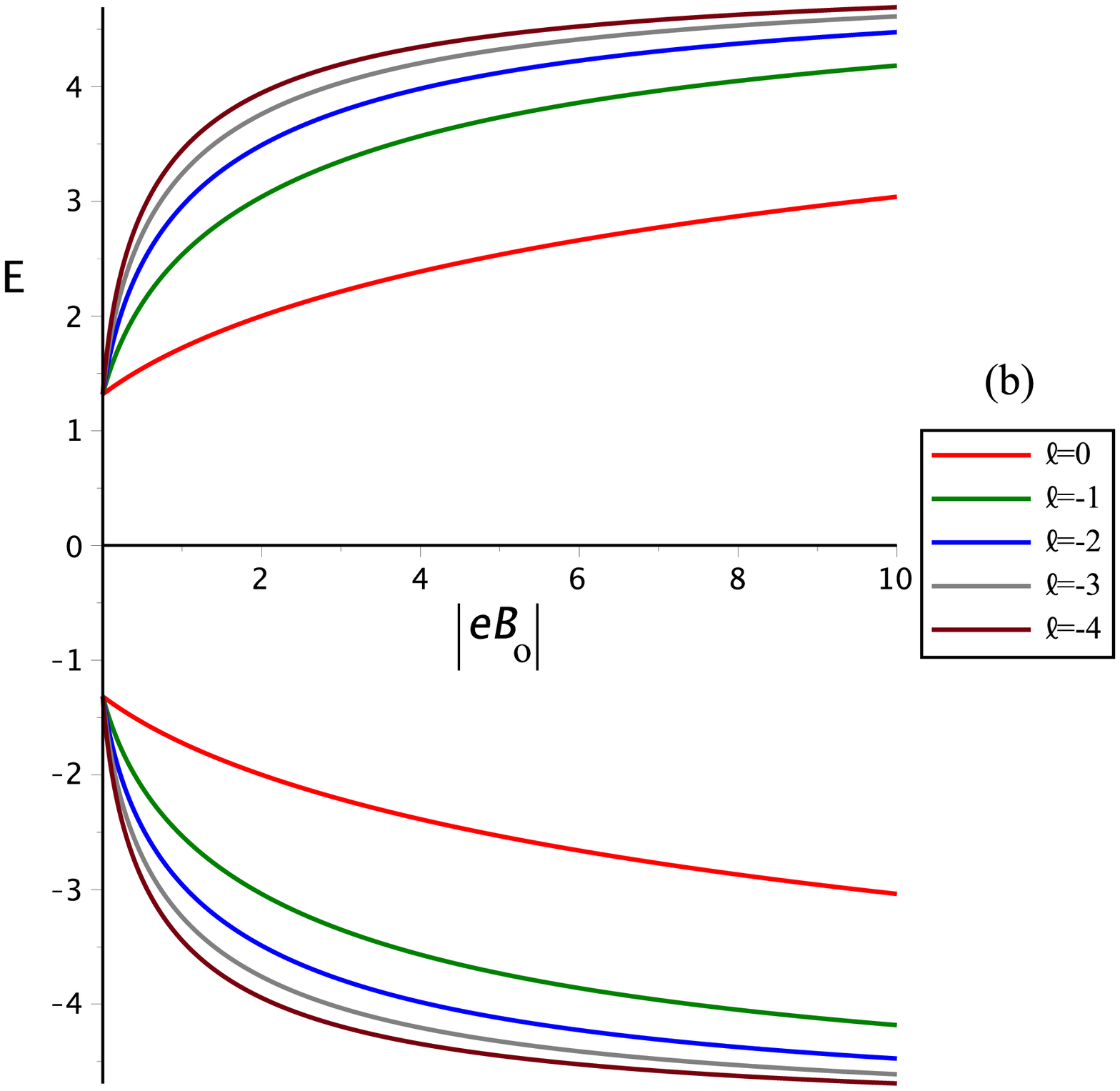} 
\caption{\small 
{ The energy levels of (\ref{e25}), using $\alpha =1/4$, $%
m=k_{z}=1$, so that (a) shows $E$ against $\beta =\epsilon /2E_{p}$ for $%
|eB_{\circ }|=1$, $n_{r}=2$, $\ell =0,1,3,6$, and (b) shows $E$ against $%
|eB_{\circ }|$ for $\beta =0.1$, $n_{r}=2$, $\ell =0,1,3,5,8$.}}
\label{fig5}
\end{figure}%
\begin{equation}
g_{_{0}}\left( y\right) ^{2}E^{2}-m^{2}=g_{_{1}}\left( y\right)
^{2}\,\mathcal{W}_{n_{r},\ell };\;\mathcal{W}_{n_{r},\ell }=\frac{\tilde{B}^{2}}{4}\left[ 1-%
\frac{\tilde{\ell}^{2}}{\left( n_{r}+|\tilde{\ell}|+\frac{1}{2}\right) ^{2}}%
\right] +k_{z}^{2}.  \label{e32}
\end{equation}%

The degeneracies associated with $\mathcal{W}_{n_{r},|\ell |}=\mathcal{W}_{n_{r},-|\ell|}$, for a given $n_r$ and $\ell$, are readily discussed in 
\cite{R40.1}.  Yet, we wish to observe the effects of rainbow gravity (with the rainbow functions fine tuned) on the spectroscopic structure of both KG-particles and anti-particles. Here, we exclude the rainbow function pair $g_{_{0}}\left( y\right) =1$, $g_{_{1}}\left(y\right) =\sqrt{1-\epsilon y^2}$ since $y^2=E^2/E_p^2$ would naturally cover the probe KG-particles and anti-particles, and the reported results in Figure 1 of \cite{R40.1} are good (the current fine tuning would not affect them).

\subsection{Rainbow functions $g_{_{0}}\left( y\right) =1$, $g_{_{1}}\left(
y\right) =\sqrt{1-\epsilon y}$ }

Now we set $g_{_{0}}\left( y\right) =1$ and $g_{_{1}}\left(
y\right) =\sqrt{1-\epsilon |E|/E_p}$ in equation (\ref{32}) to obtain%
\begin{equation}
E^{2}-m^{2}=\left( 1-\epsilon \frac{|E|}{E_{p}}\right) \mathcal{W}_{n_{r},\ell }\Longrightarrow E_{\pm }=\mp\beta \mathcal{W}_{n_{r},\ell }\pm \sqrt{\beta
^{2}\mathcal{W}_{n_{r},\ell }^{2}+\mathcal{W}_{n_{r},\ell }+m^{2}};\;\beta =\frac{\epsilon }{%
2E_{p}}.  \label{e33}
\end{equation}%
Moreover,  one may expand (\ref{e33}) about $\beta=0$ to obtain
\begin{equation}
   |E_{\pm}|\simeq \sqrt{\mathcal{W}_{n_{r},\ell }+m^2}-\mathcal{W}_{n_{r},\ell }\beta +O(\beta^2)<\sqrt{\mathcal{W}_{n_{r},\ell }+m^2}, \label{e34}
\end{equation}%
where $\sqrt{\mathcal{W}_{n_{r},\ell }+m^2}$ is the exact energy $|E_{\pm}|$ in no rainbow gravity. Moreover, as $\left\vert eB_{\circ }\right\vert\rightarrow \infty$, this result 
(\ref{e33})  would yield%
\begin{equation}
    \lim\limits_{|\tilde{B}|\rightarrow \infty }E_{\pm}\approx \pm\frac{1}{2\beta}=\pm\frac{E_p}{\epsilon }, \label{e35}
\end{equation}
which, again, suggests that $\epsilon\geq 1$. 

\begin{figure}[!ht]  
\centering
\includegraphics[width=0.3\textwidth]{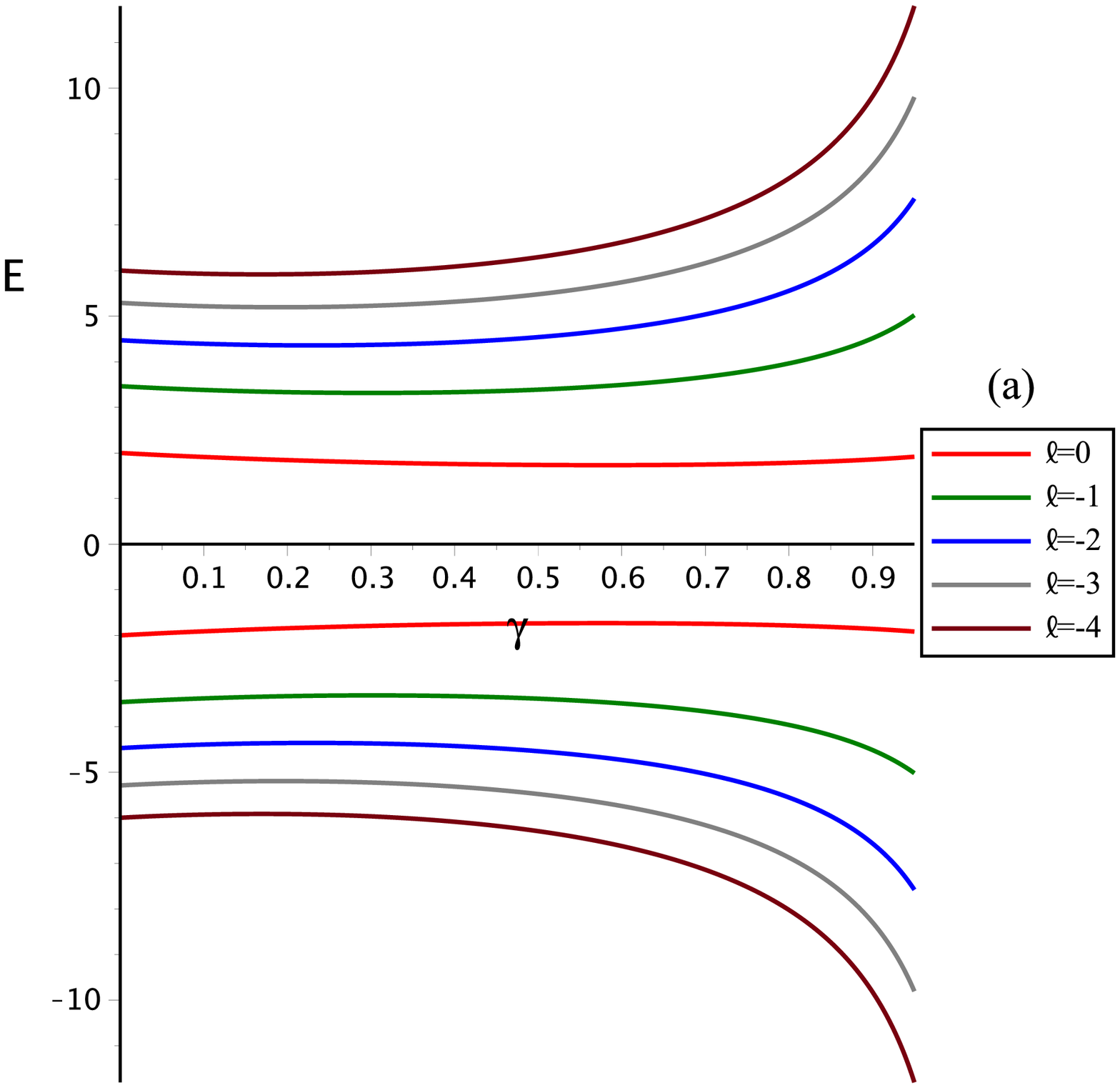}
\includegraphics[width=0.3\textwidth]{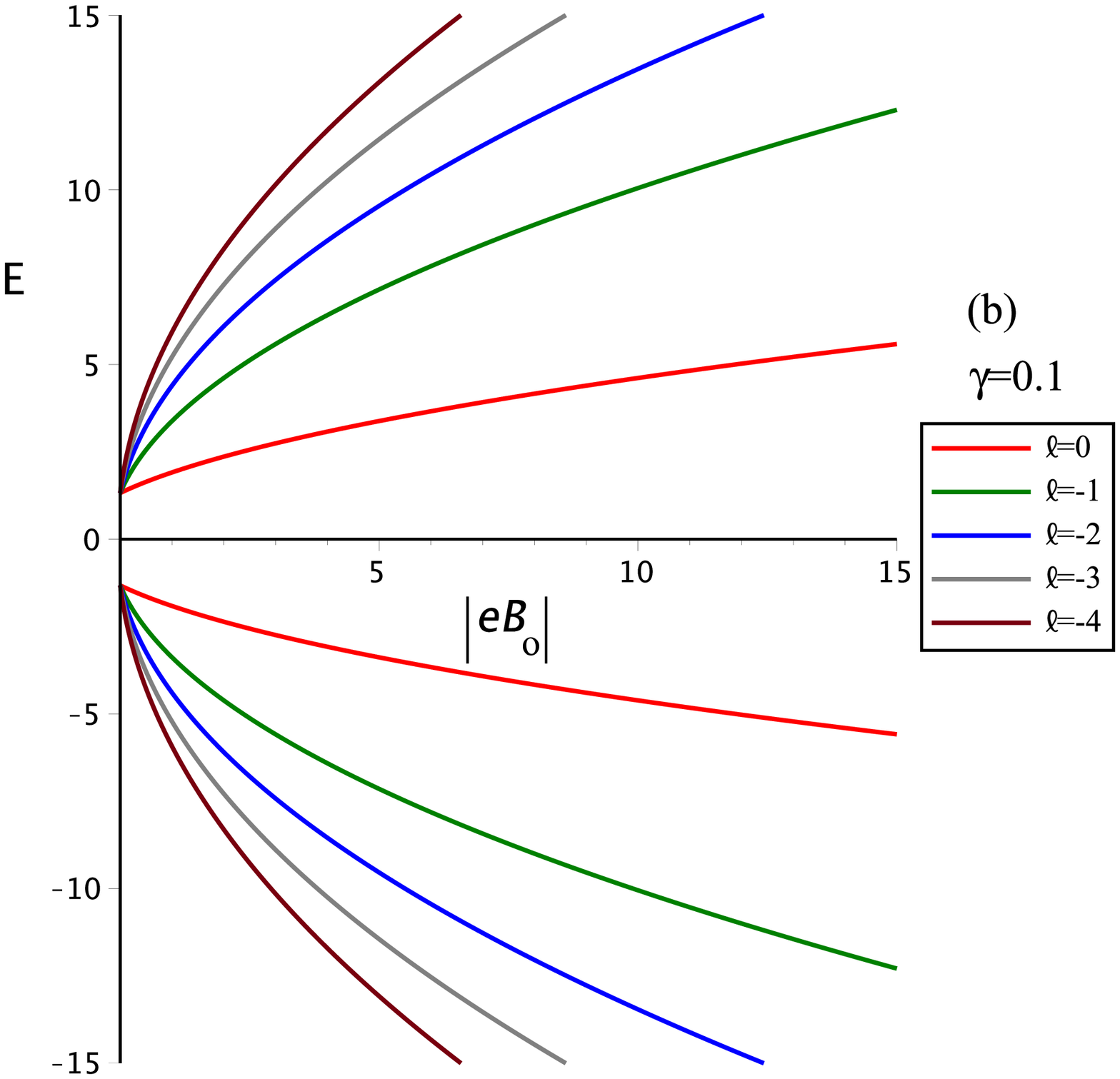}
\includegraphics[width=0.3\textwidth]{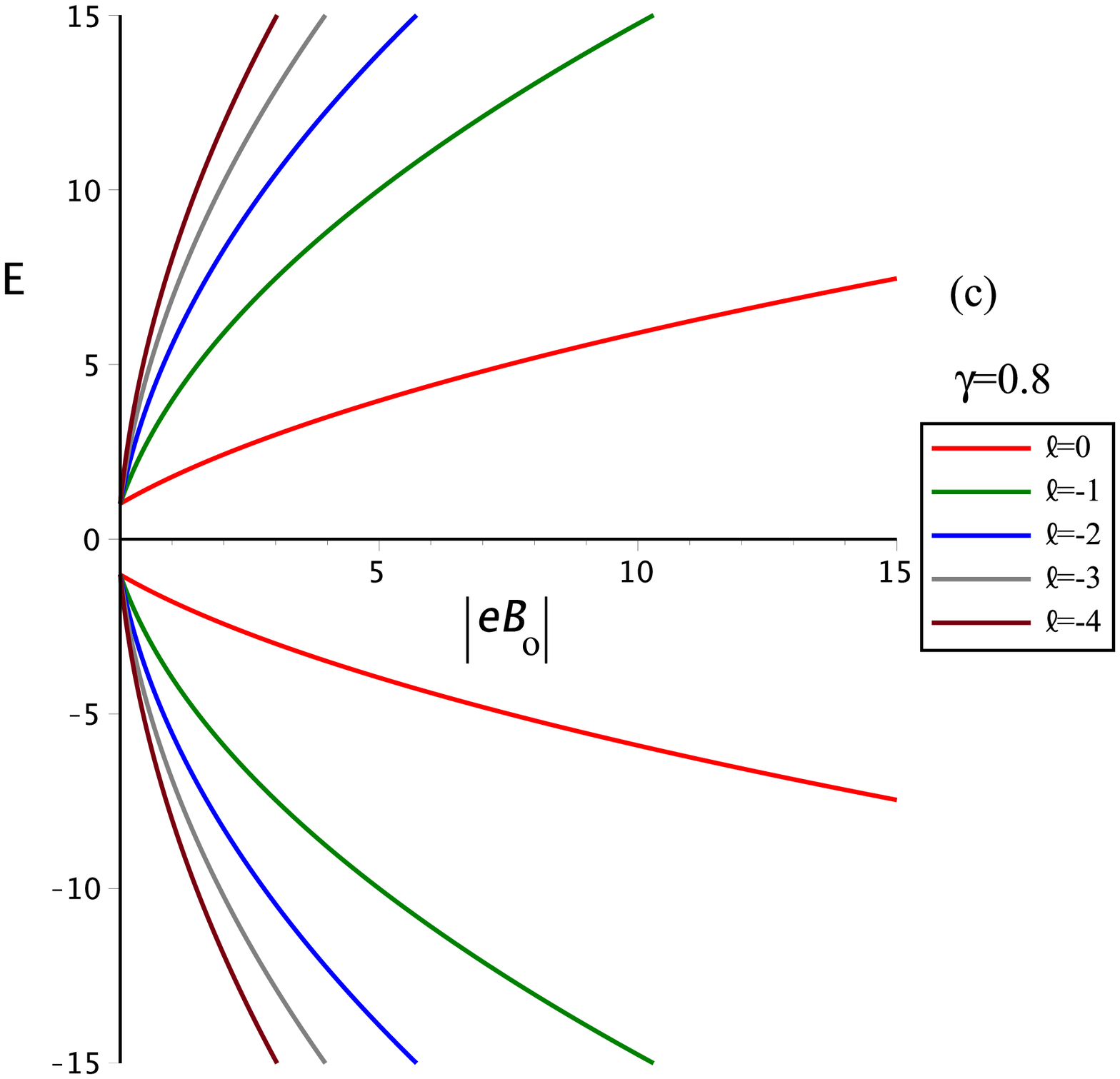}
\caption{\small 
{ The energy levels of (\ref{e36}), using $\alpha =1/4$, $%
m=k_{z}=1$, so that (a) shows $E$ against $\gamma =\epsilon m/E_{p}<1$ for $%
|eB_{\circ }|=1$, $n_{r}=2$, $\ell =0,1,3,5,8$, and (b) shows $E$ against $%
|eB_{\circ }|$ for $\gamma =0.5$, $n_{r}=2$, $\ell =0,1,3,5,8$.}}
\label{fig6}
\end{figure}%

In Figures 5(a) and (b), we plot the energy levels against $\beta =\epsilon
/2E_{p}$ and $\left\vert eB_{\circ }\right\vert $, respectively. It is obvious that the energy levels are symmetric about $E=0$. This is, in fact, what one should naturally expect from a viable approach for a KG relativistic equation. Yet, in Fig. 5(b) we clearly observe that the asymptotic tendency of the energies as $\left\vert eB_{\circ }\right\vert\rightarrow \infty $ is $E_{\pm}\approx \pm1/2\beta=\pm E_p/\epsilon$.  Moreover, Figure 2(b) shows that as $\left\vert eB_{\circ }\right\vert>>1$ the energies tend to  asymptotically converge to $E_{\pm}\approx \pm1/2\beta\approx\pm 5$, for $\beta=0.1$ used in the figure.  Whereas, it should be noted here that when $y=E/E_p$ is used ( instead of our current fine tuning $y=|E|/E_p$ )  we have observed that not only the symmetry of the energies about $E=0$ is broken but also the anti-particle energies do not secure the invariance of the Planck's energy scale $E_p$ (i.e., $|E_-|>E_p$). This is clear in Figure 2(a) and 2(b) of  \cite{R40.1}.

\subsection{Rainbow functions $g_{_{0}}\left( y\right)
=g_{_{1}}\left( y\right) =1/\left( 1-\epsilon y\right) $}

Upon the substitution of $g_{_{0}}\left( y\right) =g_{_{1}}\left( y\right)
=1/\left( 1-\epsilon y\right) $ in Eq.(\ref{e32}) we obtain%
\begin{equation}
E^{2}-\mathcal{W}_{n_{r},\ell }=\left( 1-\epsilon \frac{|E|}{E_{p}}\right)
^{2}m^{2}\Longrightarrow E_\pm=\frac{\mp\gamma m \pm \sqrt{\mathcal{W}_{n_{r},\ell }\left(
1-\gamma ^{2}\right) +m^{2}}}{1-\gamma ^{2}};\;\gamma =\frac{\epsilon m}{%
E_{p}}<1.  \label{e36}
\end{equation}%
Although an expansion about $\gamma=0$ would yield
\begin{equation}
  |E_{\pm}|\approx \sqrt{\mathcal{W}_{n_{r},\ell }+m^{2}} -m\gamma+O(\gamma^2)<|E_\pm|_{\epsilon=0}=\sqrt{\mathcal{W}_{n_{r},\ell }+m^{2}}, \label{e37}
\end{equation}
such energies fail to show any feasible convergence towards the Planck's energy scale $E_p$. Nevertheless, the symmetry of such energies about $E=0$ value is clearly documented in Figure 6(a), (b), and (c). This natural symmetry of the energies could not be achieved without the fine tuning of the rainbow functions variable (namely, $y=E/E_p$ is replaced by $y=|E|/E_p$ , see Figures 3(a)) and (b) in \cite{R40.1} for comparison).

In Figures 6(a) we plot the energy levels against $\gamma =\epsilon
m/E_{p}<1 $ to observe the rainbow gravity effect. The tendency of the energies to fly away to $\pm\infty$ and disappear from the spectrum as $\gamma\rightarrow1$ is attributed to the singularity in the energies in (\ref{e36}). In Figure 6(b) and (c) the energy levels are plotted against $\left\vert eB_{\circ
}\right\vert $, for $\gamma=0.1$ and $\gamma=0.8$, respectively.  We observe that the energy gap decreases as $\gamma$ increases, with the restriction that $\gamma<1$. %
\begin{figure}[!ht]  
\centering
\includegraphics[width=0.3\textwidth]{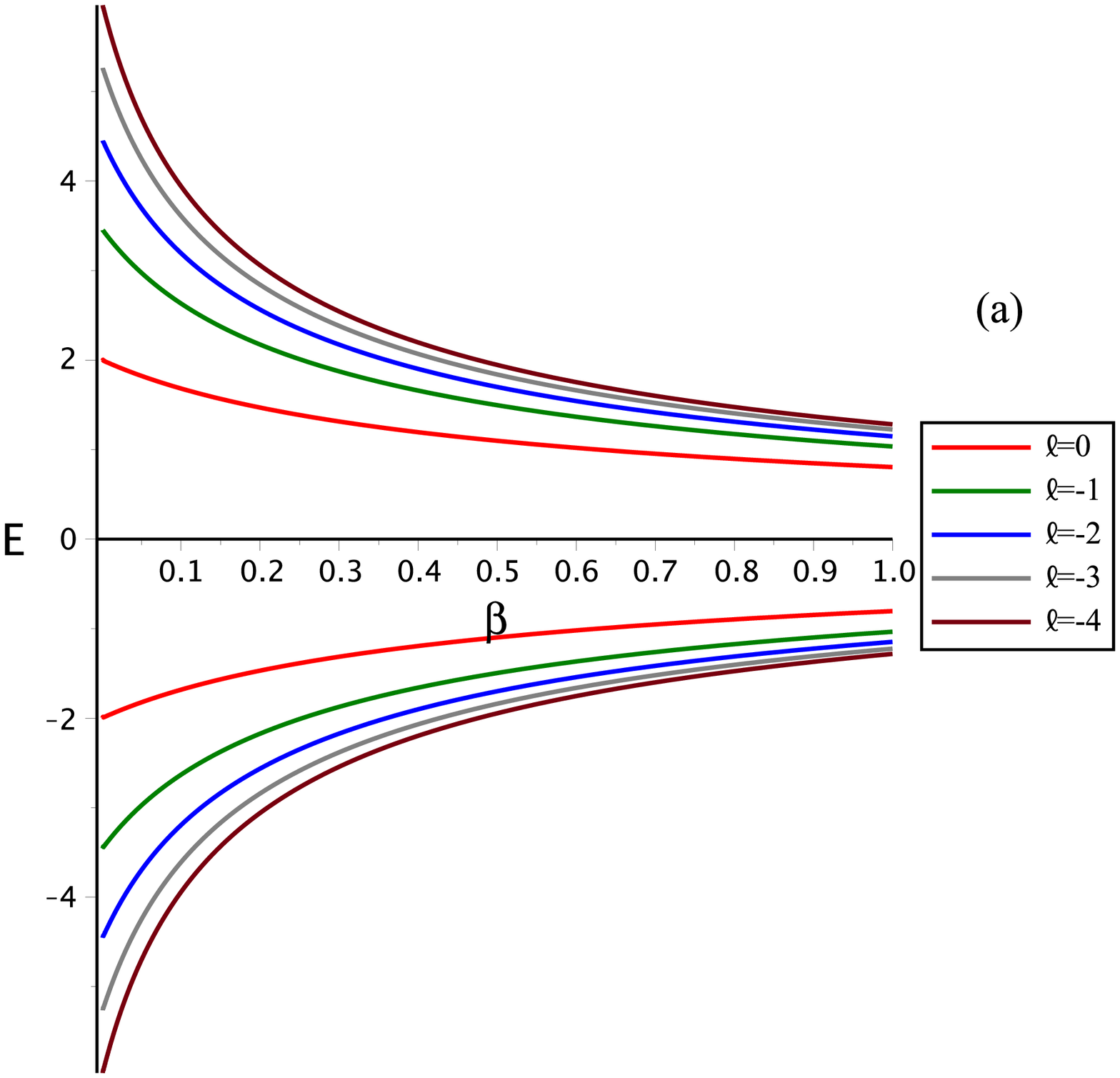}
\includegraphics[width=0.3\textwidth]{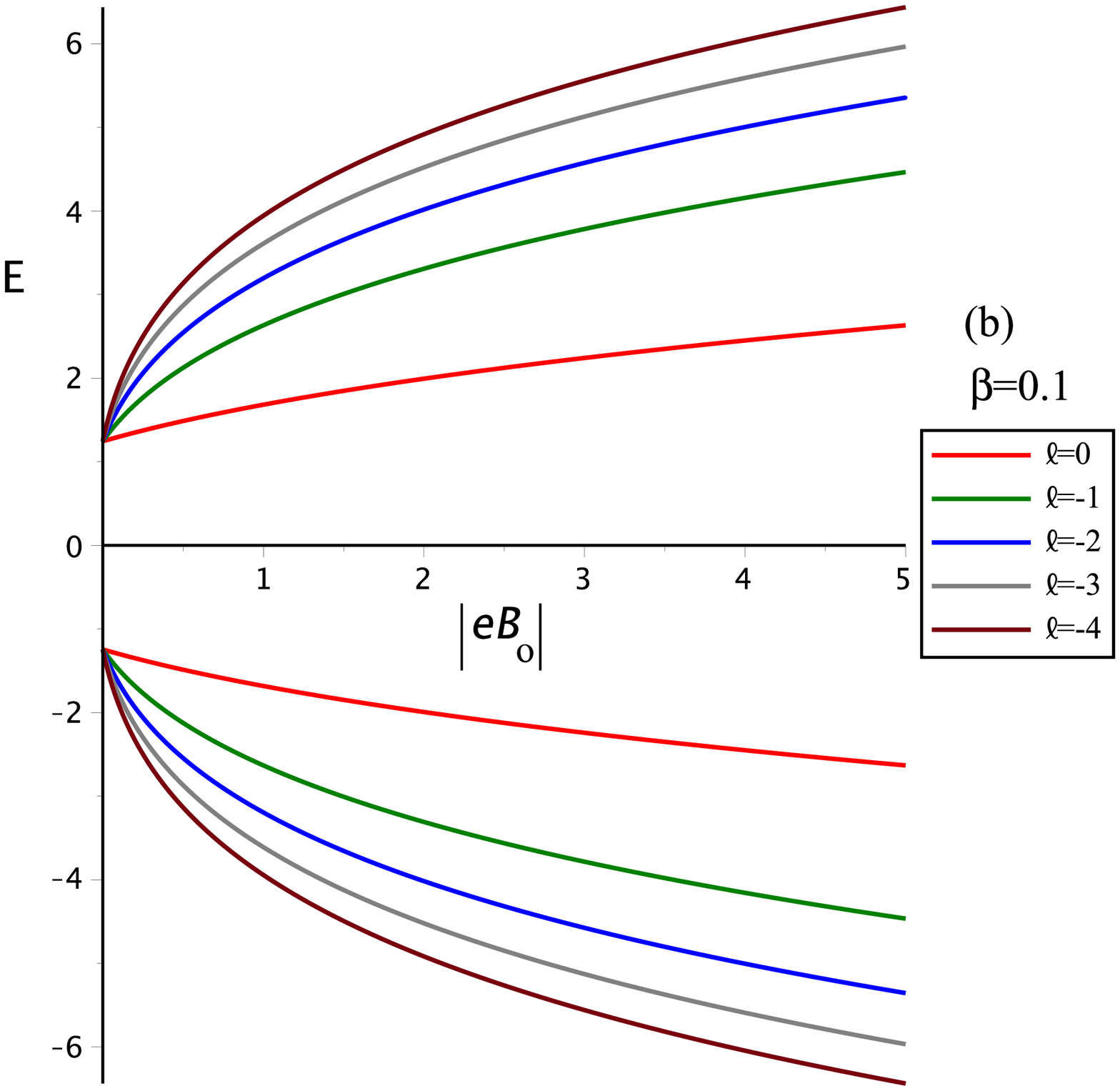} 
\includegraphics[width=0.3\textwidth]{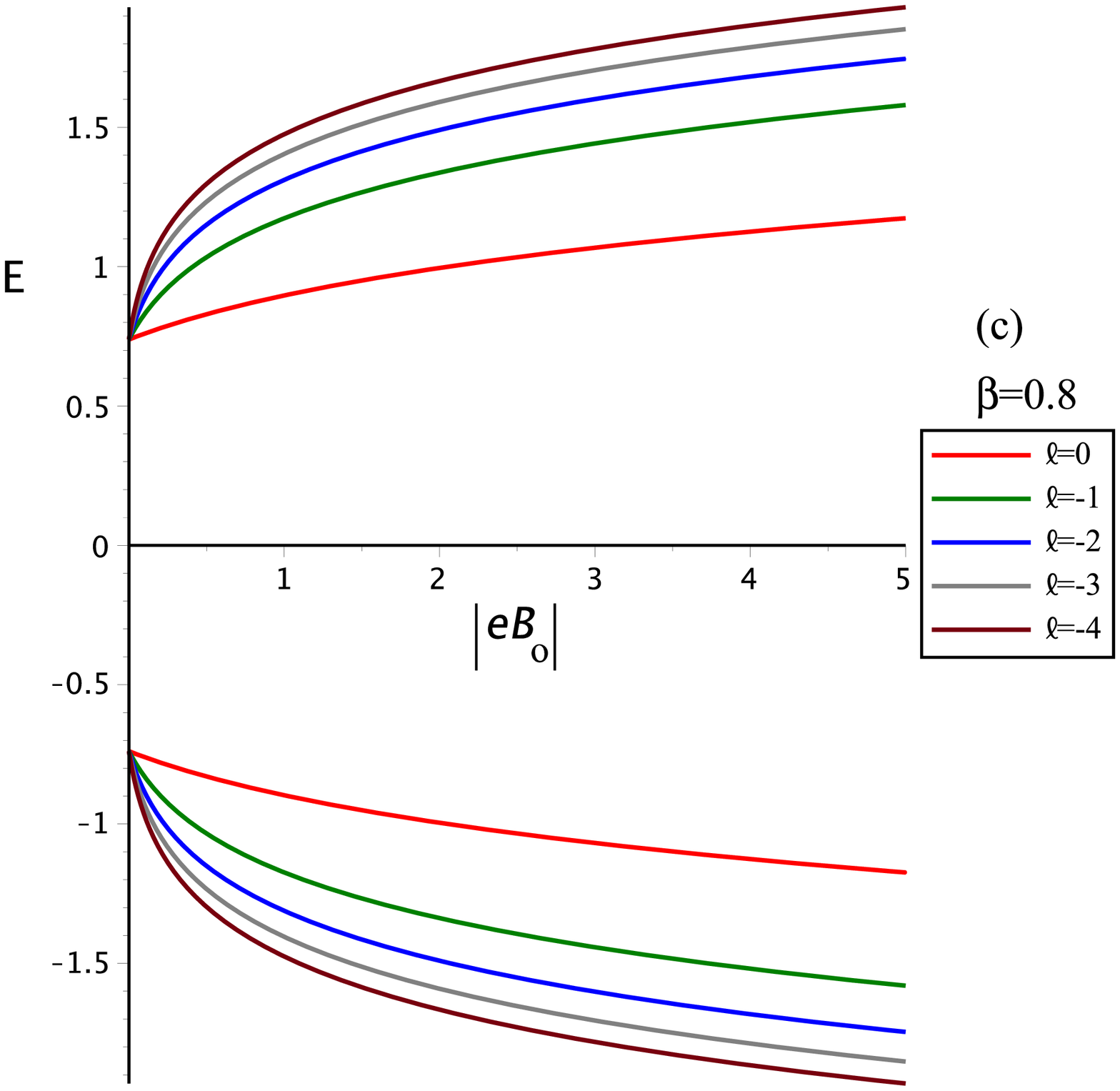}
\caption{\small 
{ The energy levels of (\ref{e38}), using $\alpha =1/4$, $%
m=k_{z}=1$, so that (a) shows $E$ against $\beta =\epsilon /2E_{p}$ for $%
|eB_{\circ }|=1$, $n_{r}=2$, $\ell =0,1,2,4$, and (b) shows $E$ against $%
|eB_{\circ }|$ for $\beta =0.1$, $n_{r}=2$, $\ell =0,1,2,3$.}}
\label{fig7}
\end{figure}%

\subsection{Rainbow functions $g_{_{0}}( y) =( e^{%
\epsilon y}-1) /\epsilon y$ and $g_{_{1}}\left(
y\right) =1$}

We now use $g_{_{0}}( y) =( e^{\epsilon y}-1)/\epsilon y$ and $g_{_{1}}\left( y\right) =1$ so that Eq.(\ref{e32}) implies%
\begin{equation}
E^{2}\left( \frac{e^{\epsilon |E|/E_{p}}-1}{\epsilon |E|/E_{p}}\right)
^{2}-m^{2}=\mathcal{W}_{n_{r},\ell }\Longrightarrow E_\pm=\pm\frac{1}{2\beta }\ln \left( 1+ 
\sqrt{4\beta ^{2}\left( \mathcal{W}_{n_{r},\ell }+m^{2}\right) }\right) ;\;\beta =\frac{\epsilon }{2E_{p}}.  \label{e38}
\end{equation}%
Where an expansion about $\beta=0$ implies
\begin{equation}
  |E_{\pm}|\approx \sqrt{\mathcal{W}_{n_{r},\ell }+m^{2}} -m\beta+O(\beta^2)<|E_\pm|_{\epsilon=0}=\sqrt{\mathcal{W}_{n_{r},\ell }+m^{2}}. \label{e39}
\end{equation}
However, because of the logarithmic nature of the result in (\ref{e38}) it is not possible to have a maximum bound  (i.e., $|E_\pm|\leq E_p$ ) for the energies. 

In Figure 7(a) we plot the energy levels against $\beta =\epsilon /2E_{p}$ . In Figures 7(b) (for $\beta=0.1$) and 7(c) (for $\beta=0.8$), we show the energy levels against $\left\vert eB_{\circ}\right\vert$. We observe that the energies are symmetric about $E=0$ value as a natural characterization of the KG-particles and anti-particles. A comparison with Figures 4(a) and 4(b) in \cite{R40.1} would emphasis that the proposed fine tuning of the rainbow functions variable is vital and necessary. Moreover,  it is clear that a comparison between 7(b) and 7(c) suggests that the energy gap narrows down as $\beta$ increases from zero. %

\section{Concluding remarks}

In a recent paper \cite{R40.1}, we have studied PDM KG-Coulomb particles in cosmic string rainbow gravity and a uniform magnetic field $\mathbf{B}=\mathbf{\nabla }\times \mathbf{A}=\frac{1}{2}B_{\circ }\,\hat{z}$  (introduced by the electromagnetic vector potential $A_{\varphi }=\frac{1}{2}B_{\circ }r$). Therein, we have found that only one rainbow functions pair (i.e., $g_{_{0}}\left( y\right) =1$, $%
g_{_{1}}\left( y\right) =\sqrt{1-\epsilon y^{2}}$; $y=E/E_p$ ) complies with the the Planck's energy scale $E_p$ invariance. Whereas, for the rainbow function pair $g_{_{0}}\left( y\right) =1$, $g_{_{1}}\left( y\right) =\sqrt{1-\epsilon y}$  (yet another member of the rainbow functions family, i.e,  $g_{_{0}}\left( y\right) =1$, $g_{_{1}}\left( y\right) =\sqrt{1-\epsilon y^{n}}$, used to describe the geometry of spacetime in loop quantum gravity \cite{R6,R61,R8,R41,R42,R421}) we found that only the KG-particle's energy $E=E_+=+|E|$ complies with rainbow gravity model but not the KG-anti-particle's energy $E=E_-=-|E|$. The only difference between the two pairs is the power of $y^n$. The former has $n=2$ and hence it covers both particles and anti-particles, whereas the later has $n=1$ that may only work for particles while the anti-particles are left unfortunate. This has, in fact, inspired our fine tuning of the rainbow function variable $y=E/E_p$ into $y=|E|/E_p$ in the current methodical proposal. 

Under such fine tuning settings, we have studied KG-oscillators in the cosmic string rainbow gravity spacetime (\ref{e4}) in a non-uniform magnetic field $%
\mathbf{B}=\mathbf{\nabla }\times \mathbf{A}=\frac{3}{2}B_{\circ }r\,\hat{z}$ (introduced by $A_{\varphi }=\frac{1}{2}B_{\circ }r^{2}$). Encouraged by the results reported for the KG-oscillators (documented in Figures 1,2,3, and 4), we have revisited KG-Coulombic particles in cosmic string rainbow gravity and uniform magnetic field (i.e., $\mathbf{B}=\frac{1}{2}B_{\circ }\,\hat{z}$) reported in \cite{R40.1}.  The results for the KG-Coulombic particles show consistency with those reported for KG-oscillators. This consistency is documented in Figures 5,6, and 7.  Namely, all energies reported for both KG-oscillators and KG-Coulombic particles are symmetric about $E=0$ value (this is the natural tendency of the energy levels for the relativistic particles and anti-particles). The loop quantum gravity pairs ($g_{_{0}}\left( y\right) =1$, $g_{_{1}}\left( y\right) =\sqrt{1-\epsilon y^2}$  and $g_{_{0}}\left( y\right) =1$, $g_{_{1}}\left( y\right) =\sqrt{1-\epsilon y}$ ), on the other hand,  have shown, beyond doubt, that they have a complete compliance with the invariance of the Planck's energy scale $E_p$. The rest of the rainbow functions, on the other hand, are not as fortunate as those of the loop quantum gravity ones. Therefore, the rainbow gravity modified relativistic energy-momentum dispersion relation should effectively be fine tuned into %
\begin{equation}
E^{2}g_{_{0}}\left( y\right) ^{2}-p^{2}c^{2}g_{_{1}}\left( y\right)
^{2}=m^{2}c^{4};\; 0\leq y=|E|/E_{p}\leq 1.  \label{e200}
\end{equation}%
where $E$ is the energy of the probe particles and anti-particles. Only under such fine tuning the modified dispersion relation (\ref{e200}) may treat relativistic probe particles and anti-particles alike.

It could be interesting to report that for the rainbow functions pair ($g_{_{0}}\left( y\right) =g_{_{1}}\left( y\right) =\left(
1-\epsilon y\right) ^{-1}$), one may observe that for $\gamma ={\epsilon m}/{E_{p}}=1$ in (\ref{e26}) and (\ref{e36}), the energy states fly away and disappear from the spectrum. A phenomenon that has been observed and reported by Mustafa and Znojil \cite{R45} for the  $\mathcal{PT}$- symmetric Schr\"{o}dinger Coulomb problem.  In the current methodical proposal, however, this phenomenon reappears as a direct effect of the rainbow gravity.

The two KG model problems  discussed in the current study should, in our opinion, form the foundation of quantum gravity as they are for the relativistic/non-relativistic quantum mechanics in the flat Minkowski spacetime (i.e., $\alpha=1$ in (\ref{e4})).  Finally, to the best of our knowledge, the current study is the first of its kind and has never been published elsewhere.

\textbf{Data availability statement:} 
The authors declare that the data supporting the findings of this study are available within the paper. 

\textbf{Declaration of interest:}
The authors declare that they have no known competing financial interests or personal relationships that could have appeared to influence the work reported in this paper.

\end{document}